\documentclass[lettersize,journal]{IEEEtran}
\usepackage{amsmath,amsfonts,amssymb}
\usepackage{graphicx}
\usepackage{algorithmicx}
\usepackage[ruled,linesnumbered]{algorithm2e}
\usepackage{algpseudocode}
\usepackage{textcomp}
\usepackage{makecell}
\usepackage{multirow,multicol}
\usepackage[colorlinks]{hyperref}
\hypersetup{citecolor=blue}
\usepackage[table]{xcolor}
\usepackage{threeparttable}
\usepackage{booktabs}
\usepackage{float}
\usepackage{balance}
\usepackage{tikz}
\usepackage{booktabs}
\usepackage{tabularx}
\usepackage{array}
\usepackage{pifont}

\newcolumntype{L}[1]{>{\raggedright\arraybackslash}p{#1}}
\newcolumntype{Y}{>{\raggedright\arraybackslash}X}
\newcolumntype{C}[1]{>{\centering\arraybackslash}m{#1}}

\usetikzlibrary{positioning,arrows.meta,calc,fit}
\usepackage[font=small,skip=2pt]{caption}

\usepackage{enumitem}
\setlist{nosep}

\begin{document}
\IEEEaftertitletext{\vspace{-10mm}}
\title{\Large Geometry-Structured Channel Reconstruction for Conventional and Fluid Antenna Systems:\\ Bayesian Inference and Fundamental Limits}

\author{Zhentian Zhang,~\IEEEmembership{Graduate Student Member,~IEEE}, Kai-Kit Wong,~\IEEEmembership{Fellow,~IEEE}, \\Kaitao Meng,~\IEEEmembership{Member,~IEEE}, David Morales-Jimenez,~\IEEEmembership{Senior Member,~IEEE}, Hao Jiang, \IEEEmembership{Senior Member,~IEEE}, \\Christos Masouros,~\IEEEmembership{Fellow,~IEEE}, Hyundong Shin, \emph{Fellow, IEEE} and Zaichen Zhang,~\IEEEmembership{Senior Member,~IEEE}
	
	\thanks{ }
	\thanks{Zhentian Zhang and Zaichen Zhang are with the National Mobile Communications Research Laboratory, Frontiers Science Center for Mobile Information Communication and Security, Southeast University, Nanjing, 210096, China. Zaichen Zhang is also with the Purple Mountain Laboratories, Nanjing 211111, China (e-mail: zhentianzhangzzt@gmail.com, zczhang@seu.edu.cn).}
	\thanks{David Morales-Jimenez is with the Department of Signal Theory, Networking and Communications, University of Granada, Granada 18071, Spain (e-mail: dmorales@ugr.es).}
	\thanks{Hao Jiang is with the National Mobile Communications Research Laboratory, Southeast University, Nanjing 210096, China (e-mail: jianghao@nuist.edu.cn).}
	\thanks{Kaitao Meng is with the Department of Electrical and Electronic Engineering, University of Manchester, Manchester, UK (email: kaitao.meng@manchester.ac.uk).}
	\thanks{Christos Masouros and Kai-Kit Wong are with the Department of Electronic and Electrical Engineering, University College London, Torrington Place, WC1E 7JE, United Kingdom. Kai-Kit Wong is also affiliated with the Department of Electronic Engineering, Kyung Hee University, Yongin-si, Gyeonggi-do 17104, Korea. (e-mail: c.masouros@ucl.ac.uk, kai-kit.wong@ucl.ac.uk).}
	\thanks{H. Shin is with the Department of Electronics and Information Convergence Engineering, Kyung Hee University, Yongin-si, Gyeonggi-do 17104, Republic of Korea (e-mail: hshin@khu.ac.kr).}
	\thanks{Corresponding authors: Kai-Kit Wong (kai-kit.wong@ucl.ac.uk)}
}



\maketitle

\begin{abstract}
Accurate channel state information (CSI) acquisition is critical for exploiting the spatial flexibility of fluid antenna systems (FASs). However, port selection and transmission optimization require CSI over a large number of candidate port positions, making direct port-wise estimation prohibitively costly in terms of pilot overhead. This paper addresses this challenge through geometry-structured channel reconstruction, which exploits the fact that the port-domain CSI can be parameterized by a small number of dominant propagation paths. We first establish fundamental mean square error (MSE) and normalized MSE (NMSE) benchmarks for both geometry-structured and unstructured channel reconstruction, providing analytical references for evaluating the intrinsic benefit of geometric modeling in conventional antenna systems and FASs. Motivated by the strong spatial correlation induced by densely distributed fluid antenna ports, we further propose a Bayesian reconstruction framework, termed geometry-structured expectation-maximization approximate message passing (GS-EM-AMP). The proposed algorithm incorporates geometric channel structure into the EM-AMP procedure and adaptively learns unknown statistical parameters from noisy observations. Numerical results demonstrate that GS-EM-AMP achieves near-bound reconstruction accuracy while maintaining strong robustness against steering-domain correlation, thereby offering an efficient and reliable solution for large-scale CSI acquisition in FASs.
\end{abstract}

\begin{IEEEkeywords}
Channel reconstruction, theoretical MSE/NMSE limit bound, approximate message passing.
\end{IEEEkeywords}
\vspace{-4mm}
\section{Introduction}

Reliable channel state information (CSI) acquisition is a fundamental prerequisite for efficient transmission, detection, estimation, and resource optimization in wireless communication systems. In millimeter-wave (mmWave) and high-frequency propagation environments, the channel often exhibits a sparse physical structure, where only a limited number of dominant scattering paths contribute significantly to the received signal. This observation has motivated the widespread use of geometry-structured channel models, also referred to as mmWave channel models with finite scatterers \cite{survey1}. Instead of estimating all channel coefficients independently, such models characterize the channel through a small set of physical parameters, including the angle-of-arrivals (AoAs), angle-of-departures (AoDs), and the corresponding path gains. When the number of dominant propagation paths is much smaller than the number of noisy channel observations, the resulting parameter dimensionality is substantially reduced \cite{Geo1, Geo4}. This geometry-induced dimensional reduction naturally leads to low-rank signal representations \cite{low_rank1,low_rank2}, thereby reducing the number of unknowns to be estimated and enabling more robust signal processing. As a result, under the same training overhead, both detection and estimation performance can be significantly improved \cite{Geo2,Geo3}.

While geometry-structured channel modeling has been extensively studied in conventional antenna systems with fixed antenna positions, its importance becomes even more pronounced in fluid antenna systems (FASs) \cite{fas-twc-21,kit_electronic}. In FASs, the antenna can be activated at one or several locations selected from a large number of candidate ports. To fully exploit the spatial flexibility of FASs, CSI over hundreds or even thousands of ports is often required for port selection, beam optimization, and multiuser coordination. Directly estimating the channel coefficients at all candidate ports would incur prohibitive pilot overhead. In contrast, under a geometry-structured model, the entire port-domain CSI can be reconstructed by estimating only a few AoAs/AoDs and their associated path gains. This makes geometry-structured modeling particularly attractive for FASs, as it can drastically reduce the channel training overhead while preserving the spatial information across a large number of ports \cite{CE5,FAS_channel 0.2,CE4,FAS_amp_jstsp,KL_fas_channel_estimation,CE6}.

Existing channel reconstruction methods for FASs can be broadly divided into two categories. The first category is based on greedy search and angle refinement. For example, \cite{CE5} considers orthogonal pilot transmission and fast Fourier transform (FFT)-based AoA/AoD detection and estimation, achieving low computational complexity and accurate channel reconstruction when sufficient spatial observations are available. However, this approach becomes less effective when the number of receiving or transmitting antennas is small, since the limited angular observations may prevent reliable FFT-based angle detection. Moreover, orthogonal pilot transmission is generally difficult to scale to multiuser or massive-connectivity scenarios. To alleviate the insufficiency of angular observations, \cite{FAS_channel 0.2} further exploits a predefined angular codebook for signal refinement. Although such greedy and codebook-based methods are computationally appealing, their performance is often sensitive to the angular resolution, codebook design, and available observation size.

The second category relies on Bayesian inference, where the essential channel parameters or statistical information are learned from the received samples. In \cite{CE4}, sparse Bayesian learning (SBL) is introduced for FAS channel estimation, where iterative learning is adopted to enhance reconstruction accuracy. More recently, \cite{FAS_amp_jstsp} extends FAS channel reconstruction to a generalized multiuser setting with non-orthogonal pilot signatures by leveraging the extremely low-complexity approximate message passing (AMP) framework \cite{EM-AMP1}. In this framework, geographical side information can be naturally incorporated into the update rules, while angular side information can be further exploited for estimation refinement. More importantly, \cite[Section~III]{FAS_amp_jstsp} reveals that the performance error floor is closely related to inaccurate modeling or inefficient learning of the true signal variance, which provides an important algorithmic principle for channel reconstruction in both conventional antenna systems and FASs. Under suitable large-system assumptions and matched statistical models, AMP-based algorithms admit tractable state-evolution analysis and can approach the minimum mean square error (MMSE) performance with complexity linear in the matrix size \cite{FAS_amp_jstsp,KL_fas_channel_estimation,CE6}, making them promising candidates for FAS channel reconstruction. In addition, \cite{KL_fas_channel_estimation} demonstrates that efficient channel reconstruction can still be achieved with only a single activated port by incorporating a more physically relevant Karhunen--Loève representation and learned AMP structures.

Despite these advances, two important issues remain insufficiently understood. First, the CSI observations in FASs can exhibit strong spatial correlation, since a large number of candidate ports are densely distributed over a limited physical aperture. Such correlation may violate the implicit independence or weak-correlation conditions under which many AMP-type algorithms are most effective. Therefore, how to incorporate geometry-structured information while maintaining robustness against strong steering-domain correlation remains an open problem. Second, most existing studies primarily evaluate channel reconstruction algorithms through numerical comparison, whereas fundamental theoretical benchmarks are still lacking. Compared with merely comparing different algorithmic designs, an analytical performance benchmark is more essential, as it characterizes the best achievable reconstruction accuracy under different channel structures, observation diversities, and training overheads. Such benchmarks are crucial for determining whether the performance loss comes from the algorithm itself or from the intrinsic limitation of the channel observation model.

Motivated by the above challenges, this paper develops both theoretical benchmarks and a robust Bayesian reconstruction framework for geometry-structured channel reconstruction in both conventional antenna systems and FASs. The main contributions are summarized as follows:
\begin{itemize}
	\item We establish fundamental mean square error (MSE) and normalized MSE (NMSE) benchmarks for channel reconstruction under both geometry-structured and geometry-unstructured models. These benchmarks provide theoretical references for evaluating channel reconstruction algorithms in conventional antenna systems and FASs.
	
	\item We propose a novel Bayesian algorithmic framework, termed geometry-structured expectation-maximization approximate message passing (GS-EM-AMP). The proposed framework incorporates geometry-structured channel information into the EM-AMP procedure and learns the unknown statistical parameters from the observations. It achieves near-bound reconstruction performance and exhibits strong robustness against the spatial correlation induced by the steering-vector structure in FASs.
\end{itemize}
\par\indent\textit{Notations:}
Bold lowercase/uppercase letters denote vectors/matrices, and calligraphic letters denote sets. $\mathbb{R}$ and $\mathbb{C}$ denote the real and complex fields. For $\mathbf{a}$ and $\mathbf{A}$, $[\mathbf{a}]_n$ and $[\mathbf{A}]_{m,n}$ denote their entries. $(\cdot)^T$, $(\cdot)^H$, $(\cdot)^*$, $\Re\{\cdot\}$, $\Im\{\cdot\}$, $\operatorname{vec}(\cdot)$, $\operatorname{tr}(\cdot)$, $\operatorname{blkdiag}(\cdot)$, $\mathbb{E}[\cdot]$, and $\operatorname{Cov}(\cdot)$ denote transpose, Hermitian transpose, conjugate, real/imaginary parts, vectorization, trace, block-diagonal concatenation, expectation, and covariance. $\otimes$, $\succeq$, and $\triangleq$ denote the Kronecker product, positive-semidefinite ordering, and definition. $\mathbf{I}$ and $\mathbf{0}$ denote identity and all-zero matrices/vectors. $|\cdot|$, $|\mathcal{A}|$, $\|\cdot\|_2$, and $\|\cdot\|_F$ denote modulus, set cardinality, Euclidean norm, and Frobenius norm. The scalar Gaussian PDFs are written as $\mathcal{CN}(x;\mu,\phi)=\frac{1}{\pi\phi}e^{-{|x-\mu|^2}/{\phi}}$ and $\mathcal{N}(x;\mu,\phi)=\frac{1}{\sqrt{2\pi\phi}}e^{-{(x-\mu)^2}/{2\phi}}$ and without the first argument, they denote the corresponding distributions.

\vspace{-4mm}
\section{Geometrical Channel and Signal Model}\label{system model}

\subsection{Signal Model}

\paragraph{Channel Training Configurations}
We consider a random-access channel training scenario, where only $K_a$ out of $K$ potential pilots/signatures are transmitted. The formulation is applicable to either uplink or downlink training and to fluid-antenna-equipped transceivers. For notational clarity, identical fluid antenna configurations are assumed, where a fluid antenna of length $W$ contains $N_o$ ports to be activated, and the channel responses at these ports are simultaneously estimated from a $G$-snapshot pilot/signature codeword. Let $\boldsymbol{a}_k\in \mathbb{C}^{G\times 1}$ denote the $k$-th user pilot, and let the BS store all potential codewords in the pilot codebook $\mathbf{A}=\left[\mathbf{a}_1,\ldots,\mathbf{a}_K\right]\in \mathbb{C}^{G\times K}$. Without loss of generality, each pilot codeword is normalized to unit power, i.e., $\|\boldsymbol{a}_k\|_2^2=1$. Due to the sparse activation of pilot/signature codewords, the receiver needs to identify which codewords are active, which corresponds to the activity detection (AD) problem.

Neglecting transmission asynchronism, the superimposed received signal from all $K$ potential sources can be written as
\begin{equation}\label{eq:1}
	\mathbf{Y} = \sum_{k=1}^{K}\alpha_k\mathbf{a}_k\mathbf{h}_k+\mathbf{Z},
\end{equation}
where $\mathbf{Y}\in\mathbb{C}^{G\times N_o}$ is the received signal, and the constant $\alpha_k$ indicates the activity state of the $k$-th pilot. Specifically, $\alpha_k=1$ means that $\boldsymbol{a}_k$ is active, whereas $\alpha_k=0$ means that it is idle. In addition, $\mathbf{Z}$ denotes the additive white Gaussian noise (AWGN), whose entries are independent and identically distributed circularly symmetric complex Gaussian random variables with zero mean and variance $\psi$, i.e., $\mathcal{CN}\left(0,\psi\right)$. The received signal model in \eqref{eq:1} can be equivalently expressed in the following compact matrix form:
\begin{equation}\label{eq:2}
	\mathbf{Y} = \mathbf{A}\mathbf{X}+\mathbf{Z},
\end{equation}
where $\mathbf{A}$ is the pilot codebook, and $\mathbf{X}\in \mathbb{C}^{K\times N_o}$ is a row-sparse matrix. Only $K_a$ rows of $\mathbf{X}$ contain nonzero entries and therefore need to be detected and estimated. Hence, \eqref{eq:2} naturally leads to a typical compressive sensing formulation.

\paragraph{Geometry-Structured Channel Model}\label{sec.channel model}
The channel vector of the $k$-th user consists of the small-scale fading component $\mathbf{s}_k$ and the large-scale fading coefficient (LSFC) $\varsigma_k$, i.e., $\mathbf{h}_k=\sqrt{\varsigma_k}\mathbf{s}_k$. For the small-scale fading component, we adopt a geometric channel model with far-field planar wave propagation and $L_s$ finite scattering paths. Let $\sigma_{k,l}$ and $\theta_{k,l}$, $l\in \{1,\ldots,L_s\}$, denote the path strength and angle-of-arrival (AoA) of the $l$-th scattering path associated with the $k$-th user, respectively. The receiving antenna is modeled as a linear array with length $W=\frac{\lambda_{len}}{2}(M-1)$, where $\frac{\lambda_{len}}{2}$ is the half-wavelength and $M$ is a positive constant. In this work, uniform port placement is assumed, and therefore the spacing between adjacent activated ports is $\frac{W}{N_o-1}$. Accordingly, the normalized steering response vector corresponding to the $l$-th scattering path of the $k$-th channel is given by
\begin{equation}\label{eq:3}
	\mathbf{s}_{k,l}= \frac{e^{ \left(-j\frac{2\pi\left(n-1\right) W}{\left(N_o-1\right)\lambda_{len}}\cos{\theta_{k,l}}\right)}}{\sqrt{N_o}},n\in\{1,\ldots,N_o\},
\end{equation}
and the small-scale fading vector $\mathbf{s}_k$ can be written as
\begin{equation}\label{eq:4}
	\mathbf{s}_k = \sum_{l=1}^{L_s}\sigma_{k,l}\mathbf{s}_{k,l}\in \mathbb{C}^{1\times N_o},
\end{equation}
where the finite scattering path model in \eqref{eq:4} can represent either a purely non-line-of-sight (NLOS) channel or a channel containing both line-of-sight (LOS) and NLOS components, depending on the strength distribution of $\sigma_{k,l}$. For the LOS-plus-NLOS case, let $K_r$ denote the Rician factor. One path strength is given by $\sqrt{\frac{K_r}{K_r+1}}e^{j\beta_k}$, where $\beta_k$ is the arbitrary LOS phase and $\Omega$ is a scaling constant. The amplitudes of the remaining path strengths are constrained by $\sqrt{\underbrace{\sigma^2_{k,1}+\ldots+\sigma^2_{k,L_s-1}}_{L_s-1 \text{~NLOS components}}}=\sqrt{\frac{1}{K_r+1}}$. Hence, the LOS AoA has a relatively stronger path strength than the NLOS AoAs.

The LSFC is determined by the distance $d_k$ in meters between the $k$-th user and the BS through a large-scale fading function $\varsigma_k=f(d_k)$. This coefficient is usually a small-valued fraction and should be compensated by a sufficient transmission energy level. To reflect practical network constraints, we assume that the channel training service is performed within the distance range $d_{ref}\le d_k \le d_{\max}$ and the angular range $\theta_{\min}\le\theta_k \le\theta_{\max}$. The distance between the BS and any user is no smaller than the reference distance $d_{ref}$, which may be caused by the altitude of the BS or by a prescribed threshold separating the near-field and far-field regions.
\vspace{-4mm}
\subsection{Review of EM-AMP}

The generalized Bernoulli-Gaussian (BG) mixture model is adopted to describe the prior distribution of $\mathbf{X}$ in \eqref{eq:2}:
\begin{equation}\label{eq:10}
	\begin{aligned}
		&p_{\mathbf{X}}\left(x_{k,n};\lambda_k,\mu^{x}_{k,n},\phi^{x}_{k,n}\right)\\
		&=(1-\lambda_k)\delta\left(x_{k,n}\right)+\lambda_k\mathcal{CN}\left(x_{k,n};\mu^{x}_{k,n},\phi^{x}_{k,n}\right),
	\end{aligned}
\end{equation}
where the non-negative constant $\lambda_k$ denotes the activity probability of the $k$-th codeword, $\mu^{x}_{k,n}$ is the mean of the desired signal component, and $\phi^{x}_{k,n}$ is the corresponding variance. The BG mixture model in \eqref{eq:10} is sufficiently flexible for different array designs and near-field scenarios, since the path responses can remain arbitrarily randomized. Moreover, when the underlying statistical distribution is complex and requires high fitting accuracy, \eqref{eq:10} can be extended to a multi-component Gaussian mixture. For instance, as shown in \cite[Fig.~1]{EM-AMP1}, the BG model provides accurate fitting capability for complex distributions.

Let $\mathbf{q}_k=\left(\lambda_k,\mu^{x}_{k,n},\phi^{x}_{k,n},\psi\right)$ collect the prior parameters to be learned from noisy observations. In the sequel, posterior quantities inferred from noisy samples are denoted by an up-arrow hat; for example, the estimate of $a$ is denoted by $\widehat{a}$.
\paragraph{Computing Posterior Statistics via Priors $\mathbf{q}_k$}
Given $\mathbf{q}_k$, AMP first characterizes the relationship between the noisy output $y_{g,n}$ and the corresponding noise-free output $r_{g,n}=\mathbf{a}_g^{\mathrm{T}}\mathbf{x}_n$, where the noise-free matrix output is denoted by $\mathbf{R}$. Here, $\mathbf{a}_g^{\mathrm{T}}$ is the $g$-th row of $\mathbf{A}$, $\mathbf{x}_n$ is the $n$-th column of $\mathbf{X}$, and $g\in\left\{1,\ldots,G\right\},n\in\left\{1,\ldots,N_o\right\}$. The conditional PDF of $y_{g,n}$ given $r_{g,n}$ is therefore
\begin{equation}\label{eq:11}
	p_{\mathbf{Y}|\mathbf{R}}(y_{g,n}|r_{g,n};\mathbf{q})=\mathcal{CN}\left(y_{g,n};r_{g,n},\psi\right)
\end{equation}
With the conditional PDF $p_{\mathbf{Y}\mid\mathbf{R}}(y_{g,n}|r_{g,n};\mathbf{q})$, the marginal posterior distribution of the noise-free output can be calculated as
\begin{equation}\label{eq:12}
	\begin{aligned}
		&p_{\mathbf{R}|\mathbf{Y}}(r_{g,n}|\mathbf{y}_n;\widehat{\mu}_{g,n}^r,\widehat{\phi}_{g,n}^r,\mathbf{q})\\
		&\triangleq\frac{p_{\mathbf{Y}|\mathbf{R}}(y_{g,n}|r_{g,n};\mathbf{q})\mathcal{CN}\left(r_{g,n};\widehat{\mu}^r_{g,n},\widehat{\phi}_{g,n}^r\right)}{\int_{r} p_{\mathbf{Y}|\mathbf{R}}(y_{g,n}|r;\mathbf{q})\mathcal{CN}\left(r;\widehat{\mu}^r_{g,n},\widehat{\phi}_{g,n}^r\right)},
	\end{aligned}
\end{equation}
where the denominator is the normalization constant. The quantities $\widehat{\mu}_{g,n}^r$ and $\widehat{\phi}_{g,n}^r$ are updated at each iteration $t$ according to \cite[Table~I, R2-R1]{EM-AMP1}, and are computed by (A1)--(A2) in Algorithm~\ref{alg:algorithm1}, respectively. Substituting \eqref{eq:11} into the numerator of \eqref{eq:12} and applying the Gaussian product identities $\mathbb{E}\left\{\mathcal{CN}(x;a,A)\mathcal{CN}(x;b,B)\right\} =\frac{aB+bA}{A+B},\mathrm{var}\left\{\mathcal{CN}(x;a,A)\mathcal{CN}(x;b,B)\right\} =\frac{AB}{A+B}$, the posterior statistics of the noise-free output are obtained as
\begin{subequations}\label{eq:13}
	\begin{align}
		&\mathbb{E}_{\mathbf{R}|\mathbf{Y}}(r_{g,n}|\mathbf{y}_n;\widehat{\mu}_{g,n}^r,\widehat{\phi}_{g,n}^r,\mathbf{q})=\frac{\widehat{\phi}_{g,n}^r\mathbf{Y}[g,n]+\psi\widehat{\mu}_{g,n}^r}{\widehat{\phi}_{g,n}^r+\psi},\label{eq:13a}\\
		&\mathrm{var}_{\mathbf{R}|\mathbf{Y}}(r_{g,n}|\mathbf{y}_n;\widehat{\mu}_{g,n}^r,\widehat{\phi}_{g,n}^r,\mathbf{q})=\frac{\widehat{\phi}_{g,n}^{r}\psi}{\widehat{\phi}_{g,n}^{r}+\psi},\label{eq:13b}
	\end{align}
\end{subequations}
where $\widetilde{\mu}^r_{g,n}$ and $\widetilde{\phi}^r_{g,n}$ are used to denote \eqref{eq:13a} and \eqref{eq:13b}, respectively. Next, AMP approximates the true marginal posterior distribution of $x_{k,n}$ as
\begin{equation}\label{eq:14}
	\begin{aligned}
		&p_{\mathbf{X}|\mathbf{Y}}(x_{k,n}|\mathbf{y}_n;\widehat{\mu}^{x}_{k,n},\widehat{\phi}^{x}_{k,n},\mathbf{q}_k) \\
		&\triangleq \frac{p_\mathbf{x}(x_{k,n};\mathbf{q}_k)\mathcal{CN}(x_{k,n};\widehat{\mu}^{x}_{k,n},\widehat{\phi}^{x}_{k,n})}{\underbrace{\int_{x} p_\mathbf{x}(x;\mathbf{q}_k)\mathcal{CN}(x;\widehat{\mu}^{x}_{k,n},\widehat{\phi}^{x}_{k,n})}_{\zeta_{k,n}}},
	\end{aligned}
\end{equation}
where $\widehat{\mu}^{x}_{k,n}$ and $\widehat{\phi}^{x}_{k,n}$ vary across iterations according to \cite[Table~I, R8-R7]{EM-AMP1}, and are computed by (A7)--(A8) in Algorithm~\ref{alg:algorithm1}, respectively. The denominator in \eqref{eq:14} is denoted by
\begin{subequations}\label{eq:15}
	\begin{align}
		\zeta_{k,n}&=\int_{x} p_\mathbf{x}(x;\mathbf{q}_k)\mathcal{CN}(x;\widehat{\mu}^{x}_{k,n},\widehat{\phi}^{x}_{k,n})\label{eq:15a}\\
		&=(1-\lambda_k)\mathcal{CN}(0;\widehat{\mu}^x_{k,n},\widehat{\phi}^x_{k.n})+\label{eq:15b}\\ &\lambda_k\mathcal{CN}(0;\widehat{\mu}^x_{k,n}-\mu^x_{k,n},\widehat{\phi}^x_{k,n}+\phi^x_{k,n}) \nonumber
	\end{align}
\end{subequations}

Substituting \eqref{eq:10} into \eqref{eq:15a} and using the Gaussian-PDF multiplication rule $\mathcal{CN}(x;a,A)\mathcal{CN}(x;b,B)=\mathcal{CN}(x;\frac{a/A+b/B}{1/A+1/B},\frac{1}{1/A+1/B})\mathcal{CN}(0;a-b,A+B)$, \eqref{eq:15b} is obtained. The purpose of this manipulation is to transform \eqref{eq:14} into a BG structure similar to \eqref{eq:10}, from which the posterior statistics of the target signal can be derived. Specifically, by substituting \eqref{eq:10} and \eqref{eq:15b} into \eqref{eq:14} and repeatedly applying the Gaussian-PDF multiplication rule, we obtain
\begin{equation}\label{eq:16}
	\begin{aligned}
		&p_{\mathbf{X}|\mathbf{Y}}(x_{k,n}|\mathbf{y}_n;\widehat{\mu}^{x}_{k,n},\widehat{\phi}^{x}_{k,n},\mathbf{q}_k)\\
		&\triangleq 
		(1-\pi_{k,n})\delta\left(x_{k,n}\right)+\pi_{k,n}\mathcal{CN}\left(x_{k,n};\gamma_{k,n},\nu_{k,n}\right),
	\end{aligned}
\end{equation}
where the parameters $\pi_{k,n}$, $\gamma_{k,n}$, and $\nu_{k,n}$ are given by
\begin{subequations}\label{eq:17}
	\begin{align}
		\gamma_{k,n}&\triangleq  \frac{\widehat{\mu}^x_{k,n}/\widehat{\phi}^x_{k,n}+\mu^x_{k,n}/\phi^x_{k,n}}{1/\widehat{\phi}^x_{k,n}+1/\phi^x_{k,n}},\label{eq:17a}\\
		\nu_{k,n}&\triangleq \frac{1}{1/\widehat{\phi}^x_{k,n}+1/\phi^x_{k,n}},\label{eq:17b}\\
		\beta_{k,n}&\triangleq \lambda_k\mathcal{CN}\left(\widehat{\mu}^x_{k,n};\mu^x_{k,n},\widehat{\phi}^x_{k,n}+\phi^x_{k,n}\right),\label{eq:17c}\\
		\pi_{k,n}&\triangleq\frac{1}{1+\left(\frac{\beta_{k,n}}{(1-\lambda_k)\mathcal{CN}(0;\widehat{\mu}^x_{k,n},\widehat{\phi}^x_{k,n})}\right)^{-1}},\label{eq:17d}
	\end{align}
\end{subequations}
where the support probability $0\le\pi_{k,n}\le1$ represents the likelihood that $x_{k,n}\neq 0$, i.e., the likelihood that the $k$-th pilot codeword is active at the $n$-th antenna. Since the codeword activity is shared across all antennas, the activity likelihood of the $k$-th codeword is determined as $\lambda_k=\frac{1}{N_o}\sum_{n=1}^{N_o}\pi_{k,n}$. Based on \eqref{eq:16}, the posterior statistics of the target signal are
\begin{subequations}\label{eq:18}
	\begin{align}
		&\mathbb{E}_{\mathbf{X}|\mathbf{Y}}\left(x_{k,n}|\mathbf{y}_n;\widehat{\mu}^x_{k,n},\widehat{\phi}^x_{k,n},\mathbf{q}_k\right)=\pi_{k,n}\gamma_{k,n},\label{eq:18a}\\
		&\mathrm{Var}_{\mathbf{X}|\mathbf{Y}}\left(x_{k,n}|\mathbf{y}_n;\widehat{\mu}^x_{k,n},\widehat{\phi}^x_{k,n},\mathbf{q}_k\right)\label{eq:18b} \\
		&= \pi_{k,n}\left(\nu_{k,n}+|\gamma_{k,n}|^2\right)-|\pi_{k,n}\gamma_{k,n}|^2,\nonumber
	\end{align}
\end{subequations} 
where \eqref{eq:18a} gives the estimate of matrix $\mathbf{X}$. In the sequel, $\widetilde{x}_{k,n}$ and $\widetilde{\phi}^x_{k,n}$ are used to denote \eqref{eq:18a} and \eqref{eq:18b}, respectively.
\begin{algorithm}[] 
	\small
	\caption{Baseline, Conventional EM-AMP \cite{EM-AMP1}}
	\label{alg:algorithm1}
	\KwIn{$\mathbf{Y}$, $\mathbf{A}$, $K$, $N_o$, $G$,  $\psi$, $T_{\max}$}
	
	\textbf{Initialize:}\\
	
	$\forall k:\lambda_k^1=\frac{G}{K}\max_{a>0}\frac{1-\frac{2K}{G}[(1+a^2)\Phi(-a)-a\mathcal{N}(a;0,1)]}{1+a^2-2[(1+a^2)\Phi(-a)-a\mathcal{N}(a;0,1)]}$\hfill \text{(I1)} \\
	
	$\forall k,n:\phi^{x,1}_{k,n}=\frac{\sum_{g=1}^{G}\left|\mathbf{Y}[g,n]\right|^{2}-M\sigma_{n}^{2}}{\sum_{g=1}^{G}\sum_{k=1}^{K}\lvert \mathbf{A}[g,k]\rvert^{2}\lambda_{k}^1}$, $\mu^{x,1}_{k,n}=0$\hfill \text{(I2)} \\
	
	$\forall k,n:\widetilde{x}^1_{k,n}=\int_{x}^{\infty}xp_{\mathbf{X}}\left(x;\lambda^1_k,\mu^{x,1}_{k,n},\phi^{x,1}_{k,n}\right)\mathrm{d}x$\hfill \text{(I3)} \\
	
	$\forall k,n:\widetilde{\phi}^1_{k,n}=\int_{x}^{\infty}\lvert x-\widetilde{x}^1_{k,n}\rvert^2 p_{\mathbf{X}}\left(x;\lambda^1_k,\mu^{x,1}_{k,n},\phi^{x,1}_{k,n}\right)\mathrm{d}x$\hfill \text{(I4)} \\
	
	$\forall g,n:\widehat{s}^0_{g,n}=0$\hfill \text{(I5)}
	
	\ForEach{$t={1,2},\cdots,T_{\max}$}{
		AMP part:
		
		$\forall g,n: \widehat{\phi}^{r,t}_{g,n}=\sum_{k=1}^{K}\lvert \mathbf{A}[g,k]\rvert^2 \phi^{x,t}_{k,n}$\hfill \text{(A1)}
		
		$\forall g,n: \widehat{\mu}^{r,t}_{g,n}=\sum_{k=1}^{K}\mathbf{A}[g,k]\widetilde{x}^t_{k,n}-\widehat{\phi}^{r,t}_{g,n}\widehat{s}^{t-1}_{g,n}$\hfill \text{(A2)}
		
		$\forall g,n: \widetilde{\mu}^{r,t}_{g,n}=\frac{\widehat{\phi}_{g,n}^{r,t}\mathbf{Y}[g,n]+\psi\widehat{\mu}_{g,n}^{r,t}}{\widehat{\phi}_{g,n}^{r,t}+\psi}$\hfill \text{(A3)}
		
		$\forall g,n: \widetilde{\phi}^{r,t}_{g,n}=\frac{\widehat{\phi}_{g,n}^{r,t}\psi}{\widehat{\phi}_{g,n}^{r,t}+\psi}$\hfill \text{(A4)}
		
		$\forall g,n: \widehat{\phi}^{s,t}_{g,n}=\frac{\widehat{\phi}^{r,t}_{g,n}-\widetilde{\phi}^{r,t}_{g,n}}{\left(\widehat{\phi}^{r,t}_{g,n}\right)^2}$\hfill \text{(A5)}
		
		$\forall g,n:
		\widehat{s}^t_{g,n}=\frac{\widetilde{\mu}^{r,t}_{g,n}-\widehat{\mu}^{r,t}_{g,n}}{\widehat{\phi}^{r,t}_{g,n}}$\hfill \text{(A6)}
		
		$\forall k,n:
		\widehat{\phi}^{x,t}_{k,n}=\left(\sum_{g=1}^{G}\lvert \mathbf{A}[g,k] \rvert^2\widehat{\phi}^{s,t}_{g,n} \right)^{-1}$\hfill \text{(A7)}
		
		$\forall k,n:
		\widehat{\mu}^{x,t}_{k,n}=\widetilde{x}^t_{k,n}+\widehat{\phi}^{x,t}_{k,n}\sum_{g=1}^{G}\left(\mathbf{A}[g,k]\right)^*\widehat{s}^t_{g,n}$\hfill \text{(A8)}
		
		$\forall k,n:
		\gamma_{k,n}\triangleq  \frac{\widehat{\mu}^{x,t}_{k,n}/\widehat{\phi}^{x,t}_{k,n}+\mu^{x,t}_{k,n}/\phi^{x,t}_{k,n}}{1/\widehat{\phi}^{x,t}_{k,n}+1/\phi^{x,t}_{k,n}}$\hfill \text{(B1)}
		
		$\forall k,n:
		\nu_{k,n}\triangleq \frac{1}{1/\widehat{\phi}^{x,t}_{k,n}+1/\phi^{x,t}_{k,n}}$\hfill \text{(B2)}
		
		$\forall k,n:
		\beta_{k,n}\triangleq \lambda^t_k\mathcal{CN}\left(\widehat{\mu}^{x,t}_{k,n};\mu^{x,t}_{k,n},\widehat{\phi}^{x,t}_{k,n}+\phi^{x,t}_{k,n}\right)$\hfill \text{(B3)}
		
		$\forall k,n:
		\pi_{k,n}\triangleq\frac{1}{1+\left(\frac{\beta_{k,n}}{(1-\lambda^t_k)\mathcal{CN}(0;\widehat{\mu}^{x,t}_{k,n},\widehat{\phi}^{x,t}_{k,n})}\right)^{-1}}$\hfill \text{(B4)}
		
		$\forall k,n:
		\widetilde{\phi}^{t+1}_{k,n}=\pi_{k,n}\left(\nu_{k,n}+|\gamma_{k,n}|^2\right)-|\pi_{k,n}\gamma_{k,n}|^2$\hfill \text{(A9)}
		
		$\forall k,n:
		\widetilde{x}^{t+1}_{k,n}=\pi_{k,n}\gamma_{k,n}$\hfill \text{(A10)}
		
		EM part:
		
		$\forall k: \lambda_k^{t+1} \triangleq \frac{1}{N_o}\sum_{n=1}^{N_o}\pi_{k,n}$\hfill \text{(E1)}
		
		$\forall k,n: \mu^{x,t+1}_{k,n} \triangleq \frac{\sum_{k=1}^{K}\pi_{k,n}\gamma_{k,n}}{\lambda_k^{t+1}K}$\hfill \text{(E2)}
		
		$\forall k,n: \phi^{x,t+1}_{k,n}\triangleq\frac{\sum_{k=1}^{K}\pi_{k,n}\left(|\mu^{x,t}_{k,n}-\gamma_{k,n}|^2+\nu_{k,n}\right)}{\lambda_k^{t+1}K}$\hfill \text{(E3)}
	}
	\KwOut{AD liklihood $\lambda_{k},k\in\left\{1,\ldots,K\right\}$ and effective CE $\widetilde{\mathbf{X}}[k,n]=\widetilde{x}^{t+1}_{k,n},k\in\left\{1,\ldots,K\right\}, n\in\left\{1,\ldots,N_o\right\}$.}
\end{algorithm}
\paragraph{Learning $\mathbf{q}_k$ From Noisy Observations}
The computations from \eqref{eq:11} to \eqref{eq:17d} show that AMP requires all parameters in $\mathbf{q}_k$ as inputs. These parameters are unknown in practice and therefore need to be learned iteratively from noisy observations. The posterior-statistics computation described above can be interpreted as the \textit{E-step} of EM and the rationale behind this interpretation can be found in \cite[Eq.18--Eq.21]{EM-AMP1}. The corresponding \textit{M-step} is formulated as
\begin{equation}\label{eq:19}
	\mathbf{q}_k^{t+1}=\arg\max_{\mathbf{q}_k^t}\widehat{\mathbb{E}}\{\ln p_{\mathbf{X}}(\mathbf{X};\mathbf{q}_k)\mid\mathbf{Y};\mathbf{q}_k^t\},
\end{equation}
where $\widehat{\mathbb{E}}$ denotes the expectation taken with respect to the AMP posterior approximation, and the superscripts $\left(t\right)$ and $\left(t+1\right)$ denote the current and next iterations, respectively. One useful interpretation of \eqref{eq:19} is that $\ln p(\mathbf{X};\mathbf{q}_k)$ acts as the log-likelihood induced by the prior distribution $p_{\mathbf{X}}(\mathbf{X};\mathbf{q}_k)$, while the posterior approximation $\widehat{\mathbb{E}}\left\{\cdot\right\}$ provides the corresponding weighting over the latent variable $\mathbf{X}$.

The resulting EM updates are given by
\begin{subequations}\label{eq:20}
	\begin{align}
		&\lambda_k^{t+1} \triangleq \frac{1}{N_o}\sum_{n=1}^{N_o}\pi_{k,n},\label{eq:20a}\\
		&\psi^{t+1} \triangleq \frac{1}{GN_o}\sum_{g=1}^{G}\sum_{n=1}^{N_o}\left(|y_{g,n}-\widetilde{\mu}^r_{g,n}|^2+\widetilde{\phi}^r_{g,n}\right),\label{eq:20b}\\
		&\mu^{x,t+1}_{k,n} \triangleq \frac{\sum_{k=1}^{K}\pi_{k,n}\gamma_{k,n}}{\lambda_k^{t+1}K},\label{eq:20c}\\
		&\phi^{x,t+1}_{k,n}\triangleq\frac{1}{\lambda_k^{t+1}K}\sum_{k=1}^{K}\pi_{k,n}\left(|\mu^{x,t}_{k,n}-\gamma_{k,n}|^2+\nu_{k,n}\right)\label{eq:20d},
	\end{align}
\end{subequations}
where $\widetilde{\mu}^r_{g,n}$ and $\widetilde{\phi}^r_{g,n}$ are defined in \eqref{eq:13a} and \eqref{eq:13b}, respectively, $\gamma_{k,n}$ is calculated by \eqref{eq:17a}, and $\nu_{k,n}$ is calculated by \eqref{eq:17b}. This completes the EM-AMP review and provides the basis for developing the proposed geometrically structured inference framework.

In the sequel, the conventional EM-AMP applied to model \eqref{eq:2} is adopted as a benchmark, and its computational procedure is summarized in Algorithm~\ref{alg:algorithm1}. Specifically, steps (A3) and (A4) follow from \eqref{eq:13b} and \eqref{eq:13a}, respectively, steps (A5) and (A6) correspond to the intermediate quantities in \cite[Table~I, R5--R6]{EM-AMP1}, steps (B1)--(B4) are obtained from \eqref{eq:17a}--\eqref{eq:17d}, steps (A9) and (A10) follow from \eqref{eq:18a} and \eqref{eq:18b}, and steps (E1)--(E3) are derived from \eqref{eq:20a}--\eqref{eq:20d}. Since the estimation of the AWGN variance is not the focus of this work, the noise variance $\psi$ is treated as a known parameter for simplicity, which reduces implementation overhead without affecting the generality of the proposed analysis. The initialization steps (I1)--(I5) follow the principles described in \cite[Section~II-D]{EM-AMP1}.

\subsection{Proposed Algorithm: Exploiting Geometry Structure}

To exploit the inherent angular sparsity of the geometrical channel, we further impose a structured prior on the active component of $\mathbf{X}$. For the $k$-th user, let $L_{\max}$ denote the maximum number of candidate propagation paths, where $l=0$ corresponds to the LOS path and $l=1,\ldots,L_{\max}-1$ correspond to candidate NLOS paths. To enable adaptive path selection, we introduce binary path indicators $b_{k,l}\in\{0,1\}$, where $b_{k,l}=1$ indicates that the $l$-th path is retained, and $b_{k,l}=0$ otherwise. Then, the prior mean of the active component at the $n$-th activated port is modeled as
\begin{equation}\label{eq:mulang1}
	\mu_{k,n}^{x}\left(\mathbf b_k,\boldsymbol{\rho}_k,\boldsymbol{\theta}_k\right)
	=
	\sum_{l=0}^{L_{\max}-1} b_{k,l}\rho_{k,l}u_n(\theta_{k,l}),
\end{equation}
where $\mathbf b_k=[b_{k,0},\ldots,b_{k,L_{\max}-1}]^{\mathrm{T}}$, $\boldsymbol{\rho}_k=[\rho_{k,0},\ldots,\rho_{k,L_{\max}-1}]^{\mathrm{T}}$, $\boldsymbol{\theta}_k=[\theta_{k,0},\ldots,\theta_{k,L_{\max}-1}]^{\mathrm{T}}$, and
\begin{equation}\label{eq:mulang2}
	u_n(\theta_{k,l})
	=
	\frac{1}{\sqrt{N_o}}
	\exp\left(
	-j\frac{2\pi(n-1)W}{(N_o-1)\lambda_{len}}\cos\theta_{k,l}
	\right).
\end{equation}
Accordingly, the prior PDF of $x_{k,n}$ is rewritten as
\begin{equation}\label{eq:mulang3}
	\begin{aligned}
		&p_{\mathbf X}(x_{k,n};\mathbf q_k)\\
		&=
		\begin{cases}
			(1-\lambda_k)\delta(x_{k,n}), & x_{k,n}=0,\\[1mm]
			\lambda_k\mathcal{CN}\!\left(
			x_{k,n};
			\mu_{k,n}^{x}\left(\mathbf b_k,\boldsymbol{\rho}_k,\boldsymbol{\theta}_k\right),
			\phi_k^x
			\right), & x_{k,n}\neq 0,
		\end{cases}
	\end{aligned}
\end{equation}
where the parameter vector is now given by
\begin{equation}\label{eq:mulang4}
	\mathbf q_k=\{\lambda_k,\phi_k^x,\mathbf b_k,\boldsymbol{\rho}_k,\boldsymbol{\theta}_k\}.
\end{equation}
Different from the element-wise independent BG prior in conventional EM-AMP, the prior in \eqref{eq:mulang3} explicitly embeds the multi-path angular structure into the mean term. As a result, the deterministic variation of the effective channel across activated ports is captured by a small number of angular components.

To adaptively determine the effective number of paths, a Bernoulli prior is imposed on the path indicators:
\begin{equation}\label{eq:mulang5}
	p(\mathbf b_k)=\prod_{l=0}^{L_{\max}-1}\kappa^{b_{k,l}}(1-\kappa)^{1-b_{k,l}},
\end{equation}
where $\kappa\in(0,1/2)$ denotes the prior probability that a candidate path survives after pruning. Since $\kappa<1/2$, activating an additional path incurs a positive penalty. The angular-parameter update is therefore formulated as the following penalized MAP problem:
\begin{equation}\label{eq:mulang6}
	\begin{aligned}
		&(\mathbf b_k^{t+1},\boldsymbol{\rho}_k^{t+1},\boldsymbol{\theta}_k^{t+1})= \\
		&\arg\max_{\mathbf b_k,\boldsymbol{\rho}_k,\boldsymbol{\theta}_k}
		\sum_{n=1}^{N_o}
		\widehat{\mathrm E}\left\{
		\ln p_{\mathbf X}(x_{k,n};\mathbf q_k)
		\middle|
		\mathbf Y,\mathbf q_k^t
		\right\}
		+\ln p(\mathbf b_k).
	\end{aligned}
\end{equation}

We next derive a tractable form of \eqref{eq:mulang6}. From the AMP recursion, the posterior PDF of $x_{k,n}$ at iteration $t$ can be written as
\begin{equation}\label{eq:mulang7}
	\begin{aligned}
		&p_{\mathbf X|\mathbf Y}(x_{k,n}|\mathbf y_n;\mathbf q_k^t)
		=\\
		&
		(1-\pi_{k,n}^{t})\delta(x_{k,n})
		+
		\pi_{k,n}^{t}\mathcal{CN}(x_{k,n};\gamma_{k,n}^{t},\nu_{k,n}^{t}),
	\end{aligned}
\end{equation}
where $\pi_{k,n}^{t}$, $\gamma_{k,n}^{t}$, and $\nu_{k,n}^{t}$ denote the posterior support probability, posterior Gaussian mean, and posterior Gaussian variance, respectively. To evaluate the expected log-prior term, the integral domain is divided into a small neighborhood around the origin and its complement, i.e., $
\mathcal{B}_{\epsilon}=\{x\in\mathbb C:|x|\le \epsilon\},~
\overline{\mathcal{B}}_{\epsilon}=\mathbb{C}\setminus \mathcal{B}_{\epsilon}$.

Then, the expected log-prior term is derived as follows:
\begin{figure*}[t!]
	\normalsize
	\begin{subequations}\label{eq:mulang10}
		\begin{align}
			&J_{k,n} \nonumber\\
			&=
			\lim_{\epsilon\to 0}\int_{x_{k,n}\in\mathcal B_{\epsilon}}
			p_{\mathbf X|\mathbf Y}(x_{k,n}|\mathbf y_n;\mathbf q_k^t)
			\ln p_{\mathbf X}(x_{k,n};\mathbf q_k)\,dx_{k,n}
			+
			\lim_{\epsilon\to 0}\int_{x_{k,n}\in\overline{\mathcal B}_{\epsilon}}
			p_{\mathbf X|\mathbf Y}(x_{k,n}|\mathbf y_n;\mathbf q_k^t)
			\ln p_{\mathbf X}(x_{k,n};\mathbf q_k)\,dx_{k,n} \label{eq:mulang10a}\\
			&=
			\underbrace{
				\lim_{\epsilon\to 0}\int_{x_{k,n}\in\mathcal B_{\epsilon}}
				p_{\mathbf X|\mathbf Y}(x_{k,n}|\mathbf y_n;\mathbf q_k^t)
				\ln\!\big[(1-\lambda_k^{t+1})\delta(x_{k,n})\big]\,dx_{k,n}
			}_{\triangleq C_{k,n}, \text{irrelevant constant component}}
			\nonumber\\
			&\qquad \qquad \qquad \qquad \qquad
			+
			\lim_{\epsilon\to 0}\int_{x_{k,n}\in\overline{\mathcal B}_{\epsilon}}
			p_{\mathbf X|\mathbf Y}(x_{k,n}|\mathbf y_n;\mathbf q_k^t)
			\ln\!\left[
			\lambda_k^{t+1}\mathcal{CN}\!\left(
			x_{k,n};
			\mu_{k,n}^{x}(\mathbf b_k,\boldsymbol{\rho}_k,\boldsymbol{\theta}_k),
			\phi_k^{x,t+1}
			\right)
			\right]dx_{k,n} \label{eq:mulang10b}\\
			&=
			C_{k,n}
			+
			\lim_{\epsilon\to 0}\int_{x_{k,n}\in\overline{\mathcal B}_{\epsilon}}
			p_{\mathbf X|\mathbf Y}(x_{k,n}|\mathbf y_n;\mathbf q_k^t)
			\ln\!\left[
			\frac{\lambda_k^{t+1}}{\pi\phi_k^{x,t+1}}
			\exp\!\left(
			-\frac{
				|x_{k,n}-\mu_{k,n}^{x}(\mathbf b_k,\boldsymbol{\rho}_k,\boldsymbol{\theta}_k)|^2
			}{
				\phi_k^{x,t+1}
			}
			\right)
			\right]dx_{k,n}
			\nonumber\\
			&=
			C_{k,n}
			+
			\ln\!\left(\frac{\lambda_k^{t+1}}{\pi\phi_k^{x,t+1}}\right)
			\underbrace{
				\lim_{\epsilon\to 0}\int_{x_{k,n}\in\overline{\mathcal B}_{\epsilon}}
				p_{\mathbf X|\mathbf Y}(x_{k,n}|\mathbf y_n;\mathbf q_k^t)\,dx_{k,n}
			}_{\triangleq \pi_{k,n}^{t}}
			\nonumber\\
			&\qquad \qquad \qquad \qquad \qquad \qquad \qquad
			-
			\frac{1}{\phi_k^{x,t+1}}
			\underbrace{
				\lim_{\epsilon\to 0}\int_{x_{k,n}\in\overline{\mathcal B}_{\epsilon}}
				p_{\mathbf X|\mathbf Y}(x_{k,n}|\mathbf y_n;\mathbf q_k^t)
				\left|
				x_{k,n}-\mu_{k,n}^{x}(\mathbf b_k,\boldsymbol{\rho}_k,\boldsymbol{\theta}_k)
				\right|^2
				dx_{k,n}
			}_{\triangleq U_{k,n}}
			\label{eq:mulang10c}\\
			&=
			C_{k,n}
			+
			\pi_{k,n}^{t}\ln\left(\frac{\lambda_k^{t+1}}{\pi\phi_k^{x,t+1}}\right)
			-\frac{1}{\phi_k^{x,t+1}}\underbrace{U_{k,n}}_{\eqref{eq:mulang15}}.
			\label{eq:mulang10d}
		\end{align}
	\end{subequations}
	\hrulefill
	\vspace*{-6mm}
\end{figure*}
In \eqref{eq:mulang10}, $C_{k,n}$ collects the terms associated with the inactive component and is independent of $(\mathbf b_k,\boldsymbol{\rho}_k,\boldsymbol{\theta}_k)$. Therefore, it does not affect the angular-parameter optimization. The remaining relevant term is $U_{k,n}$, whose closed form is derived as
\begin{figure*}[t!]
	\normalsize
	\begin{subequations}\label{eq:mulang15}
		\begin{align}
			U_{k,n}
			&=
			\pi_{k,n}^{t}
			\int
			\mathcal{CN}(x_{k,n};\gamma_{k,n}^{t},\nu_{k,n}^{t})
			\left|
			x_{k,n}-\mu_{k,n}^{x}(\mathbf b_k,\boldsymbol{\rho}_k,\boldsymbol{\theta}_k)
			\right|^2
			dx_{k,n} \label{eq:mulang15a}\\
			&=
			\pi_{k,n}^{t}
			\int
			\mathcal{CN}(x_{k,n};\gamma_{k,n}^{t},\nu_{k,n}^{t})
			\Big(
			|x_{k,n}-\gamma_{k,n}^{t}|^2
			+
			|\gamma_{k,n}^{t}-\mu_{k,n}^{x}(\mathbf b_k,\boldsymbol{\rho}_k,\boldsymbol{\theta}_k)|^2
			\nonumber\\
			&\qquad\qquad\qquad \qquad\qquad\qquad \qquad\qquad\qquad \qquad
			+
			2\Re\{(x_{k,n}-\gamma_{k,n}^{t})^*(\gamma_{k,n}^{t}-\mu_{k,n}^{x}(\mathbf b_k,\boldsymbol{\rho}_k,\boldsymbol{\theta}_k))\}
			\Big)
			dx_{k,n} \label{eq:mulang15b}\\
			&=
			\pi_{k,n}^{t}
			\left(
			\nu_{k,n}^{t}
			+
			\left|
			\gamma_{k,n}^{t}-\mu_{k,n}^{x}(\mathbf b_k,\boldsymbol{\rho}_k,\boldsymbol{\theta}_k)
			\right|^2
			\right),
			\label{eq:mulang15c}
		\end{align}
	\end{subequations}
	\hrulefill
	\vspace*{-8mm}
\end{figure*}
where \eqref{eq:mulang15b} follows from the identity
$|x-a|^2=|x-\gamma|^2+|\gamma-a|^2+2\Re\{(x-\gamma)^*(\gamma-a)\}$,
and \eqref{eq:mulang15c} uses
$\int \mathcal{CN}(x_{k,n};\gamma_{k,n},\nu_{k,n})(x_{k,n}-\gamma_{k,n})\,dx_{k,n}=0$.

Substituting \eqref{eq:mulang15} into \eqref{eq:mulang6} and removing the constants irrelevant to $(\mathbf b_k,\boldsymbol{\rho}_k,\boldsymbol{\theta}_k)$, the M-step is equivalently converted into the following penalized weighted least-squares problem:
\begin{equation}\label{eq:mulang16}
	\begin{aligned}
		&(\mathbf b_k^{t+1},\boldsymbol{\rho}_k^{t+1},\boldsymbol{\theta}_k^{t+1})
		=\arg\min_{\mathbf b_k,\boldsymbol{\rho}_k,\boldsymbol{\theta}_k}\\
		&\left(
		\sum_{n=1}^{N_o}
		\pi_{k,n}^{t}
		\left|
		\gamma_{k,n}^{t}
		-
		\sum_{l=0}^{L_{\max}-1} b_{k,l}\rho_{k,l}u_n(\theta_{k,l})
		\right|^2
		+\tau_k^{t}\|\mathbf b_k\|_0\right),
	\end{aligned}
\end{equation}
where $\tau_k^{t}=\phi_k^{x,t+1}\ln\frac{1-\kappa}{\kappa}$ is the pruning penalty, and $\|\mathbf b_k\|_0=\sum_{l=0}^{L_{\max}-1}b_{k,l}$ denotes the number of surviving paths. The first term in \eqref{eq:mulang16} fits the posterior Gaussian means $\{\gamma_{k,n}^{t}\}_{n=1}^{N_o}$ using a small number of steering responses, while the second term penalizes redundant candidate paths.

Define $\boldsymbol{\gamma}_k^{t} =[\gamma_{k,1}^{t},\ldots,\gamma_{k,N_o}^{t}]^{\mathrm{T}}$, $\mathbf W_k^{t}=\mathrm{diag}(\pi_{k,1}^{t},\ldots,\pi_{k,N_o}^{t})$, and the steering matrix
$\mathbf U_k(\boldsymbol{\theta}_k)=\left[\mathbf u(\theta_{k,0}),\ldots,\mathbf u(\theta_{k,L_{\max}-1})\right]$. Then, \eqref{eq:mulang16} can be compactly expressed as
\begin{equation}\label{eq:mulang21}
	\begin{aligned}
		&(\mathbf b_k^{t+1},\boldsymbol{\rho}_k^{t+1},\boldsymbol{\theta}_k^{t+1})
		=\\
		&\arg\min_{\mathbf b_k,\boldsymbol{\rho}_k,\boldsymbol{\theta}_k}
		\left\|
		\left(\mathbf{W}_k^{t}\right)^{1/2}
		\left(
		\boldsymbol{\gamma}_k^{t}
		-
		\mathbf U_k(\boldsymbol{\theta}_k)\mathbf D(\mathbf b_k)\boldsymbol{\rho}_k
		\right)
		\right\|_2^2
		+\tau_k^{t}\|\mathbf b_k\|_0,
	\end{aligned}
\end{equation}
where $\mathbf D(\mathbf b_k)=\mathrm{diag}(b_{k,0},\ldots,b_{k,L_{\max}-1})$.

Let $\mathcal I_k^{t+1}=\{l:b_{k,l}^{t+1}=1\}$ denote the current active path set, and let $\mathbf U_{k,\mathcal I_k}$ denote the corresponding submatrix of $\mathbf U_k(\boldsymbol{\theta}_k)$. For fixed $\mathcal I_k$ and $\boldsymbol{\theta}_k$, the optimal path-gain vector admits the weighted least-squares solution
\begin{equation}\label{eq:mulang22}
	\boldsymbol{\rho}_{k,\mathcal I_k}^{t+1}
	=
	\left(
	\mathbf U_{k,\mathcal I_k}^{H}\mathbf W_k\mathbf U_{k,\mathcal I_k}
	\right)^{-1}
	\mathbf U_{k,\mathcal I_k}^{H}\mathbf W_k\mathbf{\gamma}_k.
\end{equation}
To refine the AoA of one candidate path while keeping the other paths fixed, define the residual excluding the $l$-th path as
\begin{equation}\label{eq:mulang23}
	\mathbf r_{k,l}^{t}= \boldsymbol{\gamma}_k^{t} -\sum_{\substack{m\in\mathcal I_k^{t+1}\\m\neq l}}\rho_{k,m}^{t}\mathbf u(\theta_{k,m}^{t}).
\end{equation}
Then, the coordinate-wise subproblem associated with the $l$-th path is given by
\begin{equation}\label{eq:mulang24}
	\min_{b_{k,l}\in\{0,1\},\,\rho_{k,l},\,\theta_{k,l}}
	\left\|
	\mathbf W_k^{1/2}
	\left(
	\mathbf r_{k,l}
	-
	b_{k,l}\rho_{k,l}\mathbf u(\theta_{k,l})
	\right)
	\right\|_2^2
	+\tau_k b_{k,l}.
\end{equation}
For notational simplicity, the latest available path estimates are used in the coordinate-descent updates, as specified in steps (P5)--(P8) of Algorithm~\ref{alg:algorithm2}. If $b_{k,l}=1$, then for a fixed $\theta_{k,l}$, the optimal gain is
\begin{equation}\label{eq:mulang25}
	\rho_{k,l}^{t+1}(\theta_{k,l})
	=
	\frac{
		\mathbf u^{\mathrm{H}}(\theta_{k,l})\mathbf W_k\mathbf r_{k,l}
	}{
		\mathbf u^{\mathrm{H}}(\theta_{k,l})\mathbf W_k\mathbf u(\theta_{k,l})
	}.
\end{equation}
Substituting \eqref{eq:mulang25} into \eqref{eq:mulang24} yields the following AoA update rule:
\begin{equation}\label{eq:mulang26}
	\theta_{k,l}^{t+1}
	=
	\arg\max_{\theta_{\min}\le\theta\le\theta_{\max}}
	\frac{
		\left|
		\mathbf u^{\mathrm{H}}(\theta)\mathbf W_k\mathbf r_{k,l}
		\right|^2
	}{
		\mathbf u^{\mathrm{H}}(\theta)\mathbf W_k\mathbf u(\theta)
	}.
\end{equation}
Since $|u_n(\theta)|^2=1/N_o$, the denominator in \eqref{eq:mulang26} satisfies
\begin{equation}\label{eq:mulang27}
	\mathbf u^{\mathrm{H}}(\theta)\mathbf W_k\mathbf u(\theta)
	=
	\frac{1}{N_o}\sum_{n=1}^{N_o}\pi_{k,n},
\end{equation}
which is independent of $\theta$. Hence, \eqref{eq:mulang26} can be simplified as
\begin{equation}\label{eq:mulang28}
	\theta_{k,l}^{t+1}
	=
	\arg\max_{\theta_{\min}\le\theta\le\theta_{\max}}
	\left|
	\mathbf u^{\mathrm{H}}(\theta)\mathbf W_k\mathbf r_{k,l}
	\right|^2.
\end{equation}

The path-pruning decision is obtained by comparing the cost reduction brought by retaining the $l$-th path with the pruning penalty. Specifically, the $l$-th path is retained if and only if
\begin{equation}\label{eq:mulang29}
	\max_{\theta_{\min}\le\theta\le\theta_{\max}}
	\frac{
		\left|
		\mathbf u^{\mathrm{H}}(\theta)\mathbf W_k\mathbf r_{k,l}
		\right|^2
	}{
		\mathbf u^{\mathrm{H}}(\theta)\mathbf W_k\mathbf u(\theta)
	}
	>
	\tau_k.
\end{equation}
Otherwise, $b_{k,l}^{t+1}=0$, and the $l$-th path is pruned. Equivalently, after updating $\theta_{k,l}^{t+1}$ and $\rho_{k,l}^{t+1}$, one may define the effective path contribution as
\begin{equation}\label{eq:mulang30}
	P_{k,l}^{t+1}
	=
	|\rho_{k,l}^{t+1}|^2
	\mathbf u^{\mathrm{H}}(\theta_{k,l}^{t+1})\mathbf W_k\mathbf u(\theta_{k,l}^{t+1}),
\end{equation}
and retain the $l$-th path if $P_{k,l}^{t+1}>\tau_k$. Finally, the structured prior mean for the next iteration is updated by superposing the surviving path components:
\begin{equation}\label{eq:mulang31}
	\mu_{k,n}^{x,t+1}
	=
	\sum_{l=0}^{L_{\max}-1}
	b_{k,l}^{t+1}\rho_{k,l}^{t+1}u_n(\theta_{k,l}^{t+1}),~ n=1,\ldots,N_o.
\end{equation}
Therefore, the multi-path angular information is incorporated through the structured mean update in \eqref{eq:mulang31}, while the effective number of specular paths is adaptively determined by the pruning rule in \eqref{eq:mulang29}.

The proposed geometry-structured EM-AMP for FAS is summarized in Algorithm~\ref{alg:algorithm2}. Unlike the element-wise independent prior in conventional EM-AMP, the proposed prior adopts a shared-parameter structure across activated ports. Hence, the codeword-level activity probability $\lambda_k$ is updated in (E1) by averaging the posterior activity probabilities $\{\pi_{k,n}\}_{n=1}^{N_o}$ under the MMV common-support assumption. Similarly, after introducing the angularly structured prior mean $\mu_{k,n}^{x}=\sum_{l} b_{k,l}\rho_{k,l}u_n(\theta_{k,l})$, the deterministic port-dependent channel variation is captured by a few propagation paths. The variance $\phi_k^{x}$ is thus treated as a user-level residual variance around the structured mean, and its update in (P1) is obtained by averaging the posterior second-order residuals over all activated ports. The structured mean update in (P5)--(P9) follows from \eqref{eq:mulang21}--\eqref{eq:mulang31}.

The angular-sparsity-aware parameter update proceeds as follows:
\begin{itemize}
	\item[-] The posterior statistics $\{\pi_{k,n}^{t},\gamma_{k,n}^{t},\nu_{k,n}^{t}\}$ are first obtained from the AMP recursion in (A1)--(A10), providing empirical information on user activity and effective channel coefficients.
	\item[-] The activity probability $\lambda_k^{t+1}$ and user-level variance $\phi_k^{x,t+1}$ are updated in (E1) and (P1), which determine the weighting matrix $\mathbf W_k^{t}$ and pruning threshold $\tau_k^{t}$.
	\item[-] For each potentially active user, the structured mean is refined by solving \eqref{eq:mulang21}. The candidate AoAs are initialized from dominant angular-spectrum peaks in (P4), and then refined via coordinate-descent updates of $\theta_{k,l}$ and $\rho_{k,l}$ in (P5)--(P7).
	\item[-] Insignificant paths are pruned according to \eqref{eq:mulang29} in (P8), and the surviving paths are superposed to form $\mu_{k,n}^{x,t+1}$ according to \eqref{eq:mulang31} in (P9), which is fed back to the next AMP iteration.
\end{itemize}

Several steps in Algorithm~\ref{alg:algorithm2} are implementation-oriented approximations of the theoretical M-step:
\begin{itemize}
	\item[-] The joint minimization in \eqref{eq:mulang21} is non-convex and combinatorial. Hence, it is implemented by sequential peak initialization followed by a finite number of coordinate-descent sweeps.
	\item[-] The continuous angular search in \eqref{eq:mulang26}--\eqref{eq:mulang29} is replaced in practice by a grid search over $\Omega_{\theta}$.
	\item[-] To reduce complexity, the angular refinement may be performed only for users satisfying $\lambda_k^{t+1}>\lambda_{\min}$, while the remaining users directly set $\mu_{k,n}^{x,t+1}=0$.
\end{itemize}

\begin{algorithm}[t!] 
	\small
	\caption{Geometry-Structured EM-AMP}
	\label{alg:algorithm2}
	\KwIn{$\mathbf{Y}$, $\mathbf{A}$, $K$, $N_o$, $G$, $\psi$, $T_{\max}$, $\phi_{\min}^{x}$, $\phi_{\max}^{x}$, $L_{\max}$, $\kappa$, $\lambda_{\min}$, $C_{\max}$, $\Omega_{\theta}$}
	
	\textbf{Initialize:}~(I1)-(I5) of Algorithm~\ref{alg:algorithm1}, where $\phi_{k,n}^{x,1}\equiv \phi_k^{x,1}$ for all $n$ and $\mu_{k,n}^{x,1}=0$.
	
	\ForEach{$t={1,2},\cdots,T_{\max}$}{
		AMP part:~(A1)-(A10) of Algorithm~\ref{alg:algorithm1}, where $\phi_{k,n}^{x,t}\equiv \phi_k^{x,t}$ for all $n$.
		
		EM part:
		
		$\forall k:\lambda_k^{t+1}\triangleq \frac{1}{N_o}\sum_{n=1}^{N_o}\pi_{k,n}$\hfill \text{(E1)}
		
		$\forall k:\phi_k^{x,t+1}\triangleq
		\frac{\sum_{n=1}^{N_o}\pi_{k,n}\left(\left|\gamma_{k,n}-\mu_{k,n}^{x,t}\right|^2+\nu_{k,n}\right)}
		{\lambda_k^{t+1}N_o}$\hfill \text{(P1)}
		
		\uIf{$\phi_k^{x,t+1}>\phi_{\max}^{x}$}{$\phi_k^{x,t+1}=\phi_{\max}^{x}$}
		\ElseIf{$\phi_k^{x,t+1}<\phi_{\min}^{x}$}{$\phi_k^{x,t+1}=\phi_{\min}^{x}$}
		
		$\forall k:
		\boldsymbol{\gamma}_k^{t}\triangleq[\gamma_{k,1},\ldots,\gamma_{k,N_o}]^{T},
		\quad
		\mathbf{W}_k^{t}\triangleq\mathrm{diag}(\pi_{k,1},\ldots,\pi_{k,N_o}),
		\quad
		\tau_k^{t}\triangleq \phi_k^{x,t+1}\ln\frac{1-\kappa}{\kappa}$\hfill \text{(P2)}
		
		\ForEach{$k\in\{1,\ldots,K\}$}{
			\uIf{$\lambda_k^{t+1}\le \lambda_{\min}$}{
				$\mathcal{I}_k^{t+1}=\emptyset$ and $\forall n:\mu_{k,n}^{x,t+1}=0$\hfill \text{(P3)}
			}
			\Else{
				Initialize at most $L_{\max}$ candidate paths by selecting the dominant peaks of
				\[
				S_k^{t}(\theta)\triangleq
				\frac{\left|\mathbf{u}^{H}(\theta)\mathbf{W}_k^{t}\boldsymbol{\gamma}_k^{t}\right|^{2}}
				{\mathbf{u}^{H}(\theta)\mathbf{W}_k^{t}\mathbf{u}(\theta)},
				\quad \theta\in\Omega_{\theta},
				\]
				and denote the candidate set by $\mathcal{I}_k^{t+1}$\hfill \text{(P4)}
				
				\ForEach{$c=1,2,\ldots,C_{\max}$}{
					\ForEach{$l\in\mathcal{I}_k^{t+1}$}{
						$\mathbf{r}_{k,l}^{t}\triangleq
						\boldsymbol{\gamma}_k^{t}-
						\sum\limits_{\substack{m\in\mathcal{I}_k^{t+1}\\m\neq l}}
						\rho_{k,m}^{t}\mathbf{u}(\theta_{k,m}^{t})$\hfill \text{(P5)}
						
						$\theta_{k,l}^{t+1}\triangleq
						\arg\max\limits_{\theta\in\Omega_{\theta}}
						\frac{\left|\mathbf{u}^{H}(\theta)\mathbf{W}_k^{t}\mathbf{r}_{k,l}^{t}\right|^{2}}
						{\mathbf{u}^{H}(\theta)\mathbf{W}_k^{t}\mathbf{u}(\theta)}$\hfill \text{(P6)}
						
						$\rho_{k,l}^{t+1}\triangleq
						\frac{\mathbf{u}^{H}(\theta_{k,l}^{t+1})\mathbf{W}_k^{t}\mathbf{r}_{k,l}^{t}}
						{\mathbf{u}^{H}(\theta_{k,l}^{t+1})\mathbf{W}_k^{t}\mathbf{u}(\theta_{k,l}^{t+1})}$\hfill \text{(P7)}
						
						\uIf{$|\rho_{k,l}^{t+1}|^{2}\mathbf{u}^{H}(\theta_{k,l}^{t+1})\mathbf{W}_k^{t}\mathbf{u}(\theta_{k,l}^{t+1})\le \tau_k^{t}$}{
							prune the $l$-th path from $\mathcal{I}_k^{t+1}$\hfill \text{(P8)}
						}
					}
				}
				
				$\forall n:\mu_{k,n}^{x,t+1}\triangleq
				\sum\limits_{l\in\mathcal{I}_k^{t+1}}
				\rho_{k,l}^{t+1}u_n(\theta_{k,l}^{t+1})$\hfill \text{(P9)}
			}
		}
	}
	\KwOut{AD likelihood $\lambda_{k},k\in\{1,\ldots,K\}$, effective CE $\widetilde{\mathbf{X}}[k,n]=\widetilde{x}^{t+1}_{k,n}$, and estimated angular set $\widehat{\Theta}_k=\{\theta_{k,l}^{t+1}:l\in\mathcal{I}_k^{t+1}\}$.}
\end{algorithm}
\subsection{Complexity Analyses}\label{sec.Complexity}

We analyze the per-iteration complexities of the baseline EM-AMP and the proposed angular-sparsity-aware EM-AMP.

\paragraph{Baseline Algorithm~\ref{alg:algorithm1}}
For conventional EM-AMP, the dominant operations are the matrix multiplications in (A1)--(A2) and (A7)--(A8), requiring $\mathcal{O}(4KGN_o)$ complexity per iteration. The EM updates of $\lambda_k$, the prior mean, and the prior variance require $\mathcal{O}(KN_o)$ and $\mathcal{O}(2KN_o)$ operations, respectively. Hence, the total per-iteration complexity is
$\mathcal{O}(4KGN_o+3KN_o)$,
which is independent of $K_a$ and thus suitable for massive-connectivity scenarios.

\paragraph{Proposed Algorithm~\ref{alg:algorithm2} Exploiting Angular Sparsity}
The AMP recursion of the proposed algorithm is identical to that of Algorithm~\ref{alg:algorithm1}, with complexity $\mathcal{O}(4KGN_o)$. The updates of $\lambda_k$ and $\phi_k^x$ in (E1) and (P1) require $\mathcal{O}(2KN_o)$. Let $N_{\theta}\triangleq |\Omega_{\theta}|$ and let $K_c\le K$ denote the number of users selected for angular refinement in (P3). For each refined user, angular-spectrum evaluation over $\Omega_{\theta}$ costs $\mathcal{O}(N_oN_{\theta})$, path initialization in (P4) costs $\mathcal{O}(L_{\max}N_oN_{\theta})$, and $C_{\max}$ coordinate-descent sweeps in (P5)--(P7) cost $\mathcal{O}(C_{\max}L_{\max}N_oN_{\theta})$. The pruning and structured-mean reconstruction in (P8)--(P9) require only lower-order $\mathcal{O}(L_{\max}N_o)$ operations.

Therefore, the angular module introduces an additional complexity of
$\mathcal{O}\!\left(K_c(C_{\max}+1)L_{\max}N_oN_{\theta}\right)$
per outer iteration. In the worst case with $K_c=K$, the total per-iteration complexity becomes
$\mathcal{O}\!\left(4KGN_o+2KN_o+K(C_{\max}+1)L_{\max}N_oN_{\theta}\right)$.
Since only potentially active users are refined and $L_{\max}$ and $C_{\max}$ are small constants in practice, the proposed algorithm incurs moderate extra complexity while exploiting the geometrical structure for both conventional and FAS channels.

\section{Theoretical Limits for Unstructured and Geometry-Structured Channel Reconstruction}
\label{sec:theory}
This section establishes theoretical benchmarks for AoA-structured channel reconstruction. The goal is to quantify the dimension reduction brought by the finite-path AoA model over the conventional unstructured row-sparse model, and to characterize the fundamental performance limit when the AoAs are unknown and must be jointly estimated with the complex path gains. The considered benchmarks serve different purposes. The oracle bounds assume the active-user set is known, and some further assume known AoAs, leaving only the path gains to be estimated and thus revealing the maximum gain from AoA structure. In contrast, the joint gain-AoA CRB treats both gains and AoAs as unknown continuous parameters, making it more relevant to the proposed practical algorithm. Since activity and path-existence indicators are discrete and do not satisfy the classical CRB regularity conditions, all CRBs are conditioned on the correct active-user set and path model order, i.e., $L_k=L_s$.

\subsection{Support-Conditioned AoA-Structured Representation}
\label{subsec:theory_repr}

Recall that the received pilot signal is given by $\mathbf{Y}=\mathbf{A}\mathbf{X}+\mathbf{Z}$, where $\mathbf{A}=[\mathbf{a}_1,\ldots,\mathbf{a}_K]\in\mathbb{C}^{G\times K}$ is the pilot codebook, $\|\mathbf{a}_k\|_2^2=1$, $\mathbf{X}\in\mathbb{C}^{K\times N_o}$ is row sparse, and the entries of $\mathbf{Z}$ are i.i.d. circularly symmetric complex Gaussian random variables with variance $\psi$, i.e., $[\mathbf{Z}]_{g,n}\sim \mathcal{CN}(0,\psi)$. Let the true active-user set be
\begin{equation}
	\mathcal{S}
	\triangleq
	\{k:\alpha_k=1\},~
	|\mathcal{S}|=K_a .
	\label{eq:th_active_set}
\end{equation}

For each active user $k\in\mathcal{S}$, the channel row is assumed to be generated by a finite number of dominant propagation paths. Since the $k$-th row of $\mathbf{X}$ is a row vector, we use its transpose in the following derivation. Define the column-form steering vector
\begin{equation}
	[\mathbf{u}(\theta)]_n
	=
	\frac{1}{\sqrt{N_o}}
	\exp\left(
	-j\frac{2\pi(n-1)W}{(N_o-1)\lambda_{\rm len}}\cos\theta
	\right),
	\label{eq:th_steering}
\end{equation}
where $W$ denotes the aperture and $\lambda_{\rm len}$ denotes the wavelength. Let $L_k$ be the true number of dominant paths of user $k$, and define the effective complex path gain as
\begin{equation}
	\rho_{k,l}
	\triangleq
	\sqrt{\varsigma_k}\sigma_{k,l}.
\end{equation}
Then the transposed channel row of user $k$ can be represented as
\begin{equation}
	\mathbf{h}_k^{\mathrm{T}}
	=
	\sum_{l=1}^{L_k}
	\rho_{k,l}\mathbf{u}(\theta_{k,l})
	=
	\mathbf{U}_k(\boldsymbol{\theta}_k)\boldsymbol{\rho}_k ,
	\label{eq:th_channel_path_model}
\end{equation}
where $
\mathbf{U}_k(\boldsymbol{\theta}_k)
\triangleq
\left[
\mathbf{u}(\theta_{k,1}),
\ldots,
\mathbf{u}(\theta_{k,L_k})
\right]
\in
\mathbb{C}^{N_o\times L_k}$, $
\boldsymbol{\rho}_k
\triangleq
[\rho_{k,1},\ldots,\rho_{k,L_k}]^{\mathrm{T}}$, and $
\boldsymbol{\theta}_k
\triangleq
[\theta_{k,1},\ldots,\theta_{k,L_k}]^{\mathrm{T}}$.

Vectorizing \eqref{eq:2} yields
\begin{equation}
	\mathbf{y}
	\triangleq
	\operatorname{vec}(\mathbf{Y})
	=
	\sum_{k\in\mathcal{S}}
	\operatorname{vec}(\mathbf{a}_k\mathbf{h}_k)
	+
	\mathbf{z},
	\label{eq:th_vec_y_1}
\end{equation}
where $
\mathbf{z}
\triangleq
\operatorname{vec}(\mathbf{Z})
\sim
\mathcal{CN}(\mathbf{0},\psi\mathbf{I}_{GN_o})$.
Using the identity $
\operatorname{vec}(\mathbf{a}_k\mathbf{h}_k)
=
\mathbf{h}_k^{\mathrm{T}}\otimes \mathbf{a}_k$, and substituting \eqref{eq:th_channel_path_model} into \eqref{eq:th_vec_y_1}, we obtain
\begin{equation}
	\mathbf{y}
	=
	\sum_{k\in\mathcal{S}}
	\left(
	\mathbf{U}_k(\boldsymbol{\theta}_k)\otimes \mathbf{a}_k
	\right)
	\boldsymbol{\rho}_k
	+
	\mathbf{z}.
	\label{eq:th_vec_y_2}
\end{equation}
Let $
\mathcal{S}=\{k_1,\ldots,k_{K_a}\},~
L_{\Sigma}
\triangleq
\sum_{k\in\mathcal{S}}L_k$, and stack all path gains as $
\boldsymbol{\rho}
\triangleq
\left[
\boldsymbol{\rho}_{k_1}^{\mathrm{T}},
\ldots,
\boldsymbol{\rho}_{k_{K_a}}^{\mathrm{T}}
\right]^{\mathrm{T}}
\in
\mathbb{C}^{L_{\Sigma}}$. Define the AoA-structured sensing matrix $\mathbf{B}_{\mathcal{S},\Theta}\in
\mathbb{C}^{GN_o\times L_{\Sigma}}$ as
\begin{equation}
	\mathbf{B}_{\mathcal{S},\Theta}
	\triangleq
	\left[
	\mathbf{U}_{k_1}(\boldsymbol{\theta}_{k_1})\otimes\mathbf{a}_{k_1},
	\ldots,
	\mathbf{U}_{k_{K_a}}(\boldsymbol{\theta}_{k_{K_a}})\otimes\mathbf{a}_{k_{K_a}}
	\right],
	\label{eq:th_B_matrix}
\end{equation}
where $
\Theta
\triangleq
\{\boldsymbol{\theta}_k\}_{k\in\mathcal{S}}$. Then the observation model can be compactly written as
\begin{equation}
	\mathbf{y}
	=
	\mathbf{B}_{\mathcal{S},\Theta}\boldsymbol{\rho}
	+
	\mathbf{z}.
	\label{eq:th_structured_linear_model}
\end{equation}
The active channel vector is defined as $
\mathbf{x}_{\mathcal{S}}
\triangleq
\left[
\mathbf{h}_{k_1}^{\mathrm{T}};
\ldots;
\mathbf{h}_{k_{K_a}}^{\mathrm{T}}
\right]
\in
\mathbb{C}^{K_aN_o} $ and can be expressed as
\begin{equation}
	\mathbf{x}_{\mathcal{S}}
	=
	\mathbf{T}_{\Theta}\boldsymbol{\rho},
	\label{eq:th_T_mapping}
\end{equation}
where $
\mathbf{T}_{\Theta}
\triangleq
\operatorname{blkdiag}
\left\{
\mathbf{U}_{k_1}(\boldsymbol{\theta}_{k_1}),
\ldots,
\mathbf{U}_{k_{K_a}}(\boldsymbol{\theta}_{k_{K_a}})
\right\}
\in
\mathbb{C}^{K_aN_o\times L_{\Sigma}}$.

Equations \eqref{eq:th_structured_linear_model} and \eqref{eq:th_T_mapping} highlight the key advantage of the AoA-structured model. Instead of estimating an arbitrary $K_aN_o$-dimensional active channel vector, the structured representation only requires the estimation of $L_{\Sigma}$ complex path gains and $L_{\Sigma}$ AoAs. When $L_k\ll N_o$, this substantially reduces the intrinsic number of unknowns and can therefore improve the achievable estimation accuracy.

\subsection{Oracle Benchmarks with Known AoAs}
\label{subsec:oracle_benchmarks}

We first consider a {\em genie-aided} setting where both the true active-user set $\mathcal{S}$ and the true AoAs $\Theta$ are known. Under this assumption, the only remaining unknown continuous parameter in \eqref{eq:th_structured_linear_model} is the complex path-gain vector $\boldsymbol{\rho}$. This benchmark answers the following question: {\em if the angular structure were perfectly known, what is the best possible channel reconstruction accuracy?}

\paragraph{Oracle MMSE Benchmark}
Assume that, conditioned on $(\mathcal{S},\Theta)$, the path-gain vector follows the complex Gaussian prior $\boldsymbol{\rho}\sim \mathcal{CN}
\left(
\bar{\boldsymbol{\rho}},
\boldsymbol{\Sigma}_{\rho}
\right)$. Since \eqref{eq:th_structured_linear_model} is a linear Gaussian model, the posterior distribution of $\boldsymbol{\rho}$ is also complex Gaussian. The oracle MMSE estimator is
\begin{equation}
	\widehat{\boldsymbol{\rho}}_{\rm oracle}
	=
	\bar{\boldsymbol{\rho}}
	+
	\boldsymbol{\Sigma}_{\rho}
	\mathbf{B}_{\mathcal{S},\Theta}^{\mathrm{H}}
	\left(
	\mathbf{B}_{\mathcal{S},\Theta}
	\boldsymbol{\Sigma}_{\rho}
	\mathbf{B}_{\mathcal{S},\Theta}^{\mathrm{H}}
	+
	\psi\mathbf{I}
	\right)^{-1}
	\left(
	\mathbf{y}
	-
	\mathbf{B}_{\mathcal{S},\Theta}\bar{\boldsymbol{\rho}}
	\right).
	\label{eq:th_oracle_mmse_estimator}
\end{equation}
Equivalently, the posterior error covariance is
\begin{equation}
	\mathbf{C}_{\rho,{\rm oracle}}
	=
	\left(
	\boldsymbol{\Sigma}_{\rho}^{-1}
	+
	\frac{1}{\psi}
	\mathbf{B}_{\mathcal{S},\Theta}^{\mathrm{H}}
	\mathbf{B}_{\mathcal{S},\Theta}
	\right)^{-1}.
	\label{eq:th_oracle_mmse_cov}
\end{equation}
Using the channel mapping in \eqref{eq:th_T_mapping}, the corresponding channel-domain oracle MMSE is
\begin{equation}
	\begin{aligned}
		{\rm MMSE}_{\rm oracle}^{(X)}(\mathcal{S},\Theta)
		&=
		\mathbb{E}
		\left[
		\left\|
		\widehat{\mathbf{x}}_{\mathcal{S},{\rm oracle}}
		-
		\mathbf{x}_{\mathcal{S}}
		\right\|_2^2
		\bigm|
		\mathcal{S},\Theta
		\right]\\
		&=
		\operatorname{tr}
		\left(
		\mathbf{T}_{\Theta}
		\mathbf{C}_{\rho,{\rm oracle}}
		\mathbf{T}_{\Theta}^{\mathrm{H}}
		\right).
	\end{aligned}
	\label{eq:th_oracle_mmse_channel}
\end{equation}
Averaging over the random support and random AoAs gives
\begin{equation}
	\overline{{\rm MMSE}}_{\rm oracle}^{(X)}
	\triangleq
	\mathbb{E}_{\mathcal{S},\Theta}
	\left[
	{\rm MMSE}_{\rm oracle}^{(X)}(\mathcal{S},\Theta)
	\right].
	\label{eq:th_average_oracle_mmse}
\end{equation}
Since additional side information cannot increase the MMSE, the oracle MMSE in \eqref{eq:th_average_oracle_mmse} serves as a genie-aided lower benchmark for non-oracle estimators that do not know $(\mathcal{S},\Theta)$.

\paragraph{Oracle CRLB for Deterministic Path Gains}
We next treat the path gains as unknown deterministic parameters. Conditioned on $(\mathcal{S},\Theta)$, the log-likelihood is
\begin{equation}
	\ln p(\mathbf{y}|\boldsymbol{\rho},\mathcal{S},\Theta)
	=
	-\frac{1}{\psi}
	\left\|
	\mathbf{y}
	-
	\mathbf{B}_{\mathcal{S},\Theta}\boldsymbol{\rho}
	\right\|_2^2
	+
	{\rm const}.
	\label{eq:th_oracle_loglikelihood}
\end{equation}
Using the standard complex CRB for a linear proper-complex Gaussian model, the Fisher information matrix for $\boldsymbol{\rho}$ is
\begin{equation}
	\mathbf{J}_{\rho,{\rm oracle}}
	=
	\frac{1}{\psi}
	\mathbf{B}_{\mathcal{S},\Theta}^{\mathrm{H}}
	\mathbf{B}_{\mathcal{S},\Theta}.
	\label{eq:th_oracle_fim}
\end{equation}
If $\mathbf{B}_{\mathcal{S},\Theta}$ has full column rank, any unbiased estimator of $\boldsymbol{\rho}$ satisfies
\begin{equation}
	\operatorname{Cov}
	\left(
	\widehat{\boldsymbol{\rho}}-\boldsymbol{\rho}
	\right)
	\succeq
	\mathbf{J}_{\rho,{\rm oracle}}^{-1}
	=
	\psi
	\left(
	\mathbf{B}_{\mathcal{S},\Theta}^{\mathrm{H}}
	\mathbf{B}_{\mathcal{S},\Theta}
	\right)^{-1}.
	\label{eq:th_oracle_gain_crlb}
\end{equation}
If $\mathbf{B}_{\mathcal{S},\Theta}$ is rank deficient, the inverse in \eqref{eq:th_oracle_gain_crlb} should be interpreted on the identifiable subspace, while the variance along unidentifiable directions is unbounded.

Since $\mathbf{x}_{\mathcal{S}}=\mathbf{T}_{\Theta}\boldsymbol{\rho}$ is a linear function of $\boldsymbol{\rho}$, the function-version CRB gives
\begin{equation}
	\operatorname{Cov}
	\left(
	\widehat{\mathbf{x}}_{\mathcal{S}}
	-
	\mathbf{x}_{\mathcal{S}}
	\right)
	\succeq
	\mathbf{T}_{\Theta}
	\mathbf{J}_{\rho,{\rm oracle}}^{-1}
	\mathbf{T}_{\Theta}^{\mathrm{H}} .
	\label{eq:th_oracle_channel_cov_crlb}
\end{equation}
Therefore, the channel-domain MSE satisfies
\begin{equation}
	\mathbb{E}
	\left[
	\left\|
	\widehat{\mathbf{x}}_{\mathcal{S}}
	-
	\mathbf{x}_{\mathcal{S}}
	\right\|_2^2
	\bigm|
	\mathcal{S},\Theta
	\right]
	\ge
	\psi
	\operatorname{tr}
	\left[
	\mathbf{T}_{\Theta}
	\left(
	\mathbf{B}_{\mathcal{S},\Theta}^{\mathrm{H}}
	\mathbf{B}_{\mathcal{S},\Theta}
	\right)^{-1}
	\mathbf{T}_{\Theta}^{\mathrm{H}}
	\right],
	\label{eq:th_structured_oracle_crlb}
\end{equation}
which is the structured oracle CRLB. It quantifies the best achievable deterministic channel reconstruction accuracy when both the active support and the true AoAs are known.

\subsection{Unstructured Oracle Benchmark}
\label{subsec:unstructured_oracle}

To isolate the benefit of the AoA structure, we compare \eqref{eq:th_structured_oracle_crlb} with the oracle CRLB of a conventional unstructured row-sparse model. In the unstructured case, the support $\mathcal{S}$ is assumed known, but no angular structure is imposed. Therefore, the active channel matrix $
\mathbf{X}_{\mathcal{S}}
\in
\mathbb{C}^{K_a\times N_o}$
contains $K_aN_o$ arbitrary complex coefficients. Let $\mathbf{A}_{\mathcal{S}}\in\mathbb{C}^{G\times K_a}$ denote the submatrix of $\mathbf{A}$ formed by the active pilot columns. The observation model is
\begin{equation}
	\mathbf{Y}
	=
	\mathbf{A}_{\mathcal{S}}\mathbf{X}_{\mathcal{S}}
	+
	\mathbf{Z}.
\end{equation}
After vectorization, we have
\begin{equation}
	\mathbf{y}
	=
	\left(
	\mathbf{I}_{N_o}\otimes \mathbf{A}_{\mathcal{S}}
	\right)
	\operatorname{vec}(\mathbf{X}_{\mathcal{S}})
	+
	\mathbf{z}.
	\label{eq:th_unstructured_vec_model}
\end{equation}
Thus, the Fisher information matrix for $\operatorname{vec}(\mathbf{X}_{\mathcal{S}})$ is
\begin{equation}
	\mathbf{J}_{X,{\rm unor}}
	=
	\frac{1}{\psi}
	\left(
	\mathbf{I}_{N_o}\otimes
	\mathbf{A}_{\mathcal{S}}^{\mathrm{H}}\mathbf{A}_{\mathcal{S}}
	\right).
\end{equation}
The corresponding unstructured oracle CRLB is
\begin{align}
	\mathbb{E}
	\left[
	\left\|
	\widehat{\mathbf{X}}_{\mathcal{S}}
	-
	\mathbf{X}_{\mathcal{S}}
	\right\|_F^2
	\bigm|
	\mathcal{S}
	\right]
	&\ge
	\psi
	\operatorname{tr}
	\left[
	\left(
	\mathbf{I}_{N_o}\otimes
	\mathbf{A}_{\mathcal{S}}^{\mathrm{H}}\mathbf{A}_{\mathcal{S}}
	\right)^{-1}
	\right]                                                        \nonumber\\
	&=
	\psi N_o
	\operatorname{tr}
	\left[
	\left(
	\mathbf{A}_{\mathcal{S}}^{\mathrm{H}}\mathbf{A}_{\mathcal{S}}
	\right)^{-1}
	\right].
	\label{eq:th_unstructured_oracle_crlb}
\end{align}

The difference between \eqref{eq:th_structured_oracle_crlb} and \eqref{eq:th_unstructured_oracle_crlb} mainly comes from the dimension reduction introduced by the AoA model. To see this more clearly, consider a favorable regime where the active pilot codewords are nearly orthogonal and the steering vectors corresponding to different dominant paths are weakly coherent. In this case,
\begin{equation}
	\mathbf{A}_{\mathcal{S}}^{\mathrm{H}}\mathbf{A}_{\mathcal{S}}
	\approx
	\mathbf{I}_{K_a},~
	\mathbf{B}_{\mathcal{S},\Theta}^{\mathrm{H}}
	\mathbf{B}_{\mathcal{S},\Theta}
	\approx
	\mathbf{I}_{L_{\Sigma}} .
\end{equation}
Then the structured oracle CRLB with known angular information behaves as
\begin{equation}
	\mathbb{E}
	\left[
	\left\|
	\widehat{\mathbf{x}}_{\mathcal{S}}
	-
	\mathbf{x}_{\mathcal{S}}
	\right\|_2^2
	\bigm|
	\mathcal{S},\Theta
	\right]
	\gtrsim
	\psi L_{\Sigma},
	\label{eq:th_structured_dof_bound}
\end{equation}
whereas the unstructured oracle CRLB behaves as
\begin{equation}
	\mathbb{E}
	\left[
	\left\|
	\widehat{\mathbf{X}}_{\mathcal{S}}
	-
	\mathbf{X}_{\mathcal{S}}
	\right\|_F^2
	\bigm|
	\mathcal{S}
	\right]
	\gtrsim
	\psi K_aN_o .
	\label{eq:th_unstructured_dof_bound}
\end{equation}
Therefore, when the AoAs are known, the AoA-structured model reduces the effective number of unknown complex parameters from $K_aN_o$ arbitrary channel coefficients to $L_{\Sigma}$ dominant-path gains.

Under the normalization $\|\mathbf{a}_k\|_2^2=1$, the pilot length $G$ does not appear as a simple multiplicative factor in the CRLB. Instead, $G$ affects the bound through the conditioning of $\mathbf{A}_{\mathcal{S}}^{\mathrm{H}}\mathbf{A}_{\mathcal{S}}$, or more generally through the mutual coherence among pilot codewords. A larger pilot dimension provides more room to design less coherent pilot sequences, which improves the conditioning of the Fisher information matrix and leads to a tighter estimation bound.

\subsection{Joint Gain-AoA CRB with Unknown AoAs}
\label{subsec:joint_crb}

The oracle CRLBs in \eqref{eq:th_structured_oracle_crlb} and \eqref{eq:th_unstructured_oracle_crlb} assume that the AoAs are perfectly known. We now remove this assumption and treat the AoAs as unknown deterministic parameters. The active set $\mathcal{S}$ and the path numbers $\{L_k\}_{k\in\mathcal{S}}$ are still assumed known, so that the CRB is applied only to continuous parameters.

\subsubsection{Real-Valued Parameterization}

Globally index all surviving paths by $q=1,\ldots,L_{\Sigma}$, and let $\mathsf{k}(q)\in\mathcal{S}$ denote the user associated with the $q$-th path. The mean of the observation can be written as
\begin{equation}
	\boldsymbol{\mu}(\boldsymbol{\xi})
	=
	\sum_{q=1}^{L_{\Sigma}}
	\rho_q\mathbf{b}_q(\theta_q),
	\label{eq:th_mu_sum}
\end{equation}
where $
\mathbf{b}_q(\theta_q)
\triangleq
\mathbf{u}(\theta_q)\otimes \mathbf{a}_{\mathsf{k}(q)}
\in
\mathbb{C}^{GN_o}$. The unknown continuous parameter vector is written in real form as $
\boldsymbol{\xi}
\triangleq
\left[
\boldsymbol{\alpha}_R^{\mathrm{T}},
\boldsymbol{\alpha}_I^{\mathrm{T}},
\boldsymbol{\theta}^{\mathrm{T}}
\right]^{\mathrm{T}}
\in
\mathbb{R}^{3L_{\Sigma}}$, where $
\boldsymbol{\rho}
=
\boldsymbol{\alpha}_R
+
j\boldsymbol{\alpha}_I,
\qquad
\boldsymbol{\theta}
=
[\theta_1,\ldots,\theta_{L_{\Sigma}}]^{\mathrm{T}}$. Therefore, the observation follows the Gaussian model $
\mathbf{y}
\sim
\mathcal{CN}
\left(
\boldsymbol{\mu}(\boldsymbol{\xi}),
\psi\mathbf{I}_{GN_o}
\right)$.

\subsubsection{Jacobian of the Mean Vector}

For the real and imaginary parts of the path gains, we have
\begin{equation}
	\frac{\partial \boldsymbol{\mu}}{\partial \alpha_{R,q}}
	=
	\mathbf{b}_q(\theta_q),~
	\frac{\partial \boldsymbol{\mu}}{\partial \alpha_{I,q}}
	=
	j\mathbf{b}_q(\theta_q).
	\label{eq:th_gain_derivatives}
\end{equation}
For the AoA parameter $\theta_q$, define
\begin{equation}
	\dot{\mathbf{b}}_q(\theta_q)
	\triangleq
	\frac{\partial \mathbf{b}_q(\theta_q)}{\partial \theta_q}
	=
	\dot{\mathbf{u}}(\theta_q)
	\otimes
	\mathbf{a}_{\mathsf{k}(q)} .
	\label{eq:th_dot_bq}
\end{equation}
Then, $\frac{\partial \boldsymbol{\mu}}{\partial \theta_q}
=
\rho_q\dot{\mathbf{b}}_q(\theta_q)$. Define $
\mathbf{B}(\boldsymbol{\theta})
\triangleq
\left[
\mathbf{b}_1(\theta_1),
\ldots,
\mathbf{b}_{L_{\Sigma}}(\theta_{L_{\Sigma}})
\right]$, and $
\mathbf{D}(\boldsymbol{\rho},\boldsymbol{\theta})
\triangleq
\left[
\rho_1\dot{\mathbf{b}}_1(\theta_1),
\ldots,
\rho_{L_{\Sigma}}\dot{\mathbf{b}}_{L_{\Sigma}}(\theta_{L_{\Sigma}})
\right]$.
The complex Jacobian of the mean with respect to the real parameter vector $\boldsymbol{\xi}$ is therefore $
\mathbf{G}_{\mu}(\boldsymbol{\xi})
=
\left[
\mathbf{B}(\boldsymbol{\theta}),
\,
j\mathbf{B}(\boldsymbol{\theta}),
\,
\mathbf{D}(\boldsymbol{\rho},\boldsymbol{\theta})
\right]$.

\subsubsection{Deterministic Fisher Information Matrix}

For a proper-complex Gaussian observation with parameter-dependent mean and parameter-independent covariance $\psi\mathbf{I}$, the Slepian--Bangs formula gives the real-valued Fisher information matrix as
\begin{equation}
	\mathbf{J}_D(\boldsymbol{\xi})
	=
	\frac{2}{\psi}
	\Re
	\left\{
	\mathbf{G}_{\mu}^{\mathrm{H}}(\boldsymbol{\xi})
	\mathbf{G}_{\mu}(\boldsymbol{\xi})
	\right\}.
	\label{eq:th_joint_fim_general}
\end{equation}
Let $
\mathbf{R}
\triangleq
\mathbf{B}^{\mathrm{H}}\mathbf{B},~
\mathbf{Q}
\triangleq
\mathbf{B}^{\mathrm{H}}\mathbf{D},~
\mathbf{P}
\triangleq
\mathbf{D}^{\mathrm{H}}\mathbf{D}$. Then \eqref{eq:th_joint_fim_general} can be written explicitly as
\begin{equation}
	\mathbf{J}_D(\boldsymbol{\xi})
	=
	\frac{2}{\psi}
	\begin{bmatrix}
		\Re\{\mathbf{R}\}
		&
		-\Im\{\mathbf{R}\}
		&
		\Re\{\mathbf{Q}\}
		\\[1mm]
		\Im\{\mathbf{R}\}
		&
		\Re\{\mathbf{R}\}
		&
		\Im\{\mathbf{Q}\}
		\\[1mm]
		\Re\{\mathbf{Q}^{\mathrm{H}}\}
		&
		-\Im\{\mathbf{Q}^{\mathrm{H}}\}
		&
		\Re\{\mathbf{P}\}
	\end{bmatrix}.
	\label{eq:th_joint_fim_block}
\end{equation}
Consequently, any unbiased estimator of $\boldsymbol{\xi}$ satisfies
\begin{equation}
	\operatorname{Cov}
	\left(
	\widehat{\boldsymbol{\xi}}
	-
	\boldsymbol{\xi}
	\right)
	\succeq
	\mathbf{J}_D^{-1}(\boldsymbol{\xi}).
	\label{eq:th_joint_crb_xi}
\end{equation}
The block structure in \eqref{eq:th_joint_fim_block} has a clear interpretation:
\begin{itemize}
	\item[-] $\mathbf{B}^{\mathrm{H}}\mathbf{B}$ determines the information about the complex path gains.
	\item[-] $\mathbf{D}^{\mathrm{H}}\mathbf{D}$ determines the information about the AoAs.
	\item[-] $\mathbf{B}^{\mathrm{H}}\mathbf{D}$ quantifies the coupling between path-gain estimation and AoA estimation.
\end{itemize}
Thus, when the AoAs are unknown, the gain estimation error and the angular estimation error are generally coupled and should not be analyzed separately.

\subsubsection{Derivative of the Steering Vector}

For the steering vector in \eqref{eq:th_steering}, define $
\beta_n
\triangleq
\frac{2\pi(n-1)W}{(N_o-1)\lambda_{\rm len}}$. Then $
[\mathbf{u}(\theta)]_n
=
\frac{1}{\sqrt{N_o}}
\exp(-j\beta_n\cos\theta)$.
Taking the derivative with respect to $\theta$ gives
\begin{equation}
	\dot{u}_n(\theta)
	=
	\frac{\partial u_n(\theta)}{\partial \theta}
	=
	j\beta_n\sin\theta\,u_n(\theta).
	\label{eq:th_steering_derivative_entry}
\end{equation}
Hence, $
\dot{\mathbf{u}}(\theta)
=
\left[
j\beta_1\sin\theta\,u_1(\theta),
\ldots,
j\beta_{N_o}\sin\theta\,u_{N_o}(\theta)
\right]^{\mathrm{T}} $, whose squared norm is
\begin{align}
	\left\|
	\dot{\mathbf{u}}(\theta)
	\right\|_2^2
	&=
	\frac{\sin^2\theta}{N_o}
	\sum_{n=1}^{N_o}
	\left(
	\frac{2\pi(n-1)W}{(N_o-1)\lambda_{\rm len}}
	\right)^2                                                        \nonumber\\
	&=
	\frac{2\pi^2W^2(2N_o-1)}
	{3(N_o-1)\lambda_{\rm len}^2}
	\sin^2\theta .
	\label{eq:th_steering_derivative_norm}
\end{align}
This expression shows that the AoA sensitivity increases with the aperture $W$ and the number of activated ports $N_o$. It also shows that angles with small $|\sin\theta|$ are more difficult to estimate, because the steering vector changes more slowly with respect to $\theta$.

When the complex path gain is also unknown, the angular information is not determined solely by $\|\dot{\mathbf{u}}(\theta)\|_2^2$. This is because the component of $\dot{\mathbf{u}}(\theta)$ parallel to $\mathbf{u}(\theta)$ can be absorbed by the unknown complex gain. Therefore, after eliminating the nuisance gain, the effective angular sensitivity is governed by the projected derivative
\begin{equation}
	\dot{\mathbf{u}}_{\perp}(\theta)
	\triangleq
	\left(
	\mathbf{I}
	-
	\mathbf{u}(\theta)\mathbf{u}^{\mathrm{H}}(\theta)
	\right)
	\dot{\mathbf{u}}(\theta).
	\label{eq:th_projected_derivative}
\end{equation}
Since $\|\mathbf{u}(\theta)\|_2=1$, we have
\begin{align}
	\left\|
	\dot{\mathbf{u}}_{\perp}(\theta)
	\right\|_2^2
	&=
	\left\|
	\dot{\mathbf{u}}(\theta)
	\right\|_2^2
	-
	\left|
	\mathbf{u}^{\mathrm{H}}(\theta)\dot{\mathbf{u}}(\theta)
	\right|^2                                                       \nonumber\\
	&=
	\frac{\pi^2W^2(N_o+1)}
	{3(N_o-1)\lambda_{\rm len}^2}
	\sin^2\theta .
	\label{eq:th_projected_derivative_norm}
\end{align}
The projected derivative in \eqref{eq:th_projected_derivative_norm} is the relevant quantity for scalar AoA CRB interpretation when the complex path gain is unknown.

\subsubsection{Approximate Scalar AoA CRB}

The exact CRB is obtained by inverting the full Fisher information matrix in \eqref{eq:th_joint_fim_block}. Nevertheless, a more intuitive scalar expression can be obtained under a weak-coupling approximation. Suppose that different paths are nearly orthogonal and that the dominant coupling for each path is between its own complex gain and its own AoA. Then, after treating the complex gain as a nuisance parameter, the AoA variance of path $q$ approximately satisfies
\begin{equation}
	\operatorname{var}(\widehat{\theta}_q)
	\gtrsim
	\frac{\psi}
	{
		2|\rho_q|^2
		\|\mathbf{a}_{\mathsf{k}(q)}\|_2^2
		\left\|
		\dot{\mathbf{u}}_{\perp}(\theta_q)
		\right\|_2^2
	} .
	\label{eq:th_scalar_aoa_crb_general}
\end{equation}
Using $\|\mathbf{a}_{\mathsf{k}(q)}\|_2^2=1$ and \eqref{eq:th_projected_derivative_norm}, this becomes
\begin{equation}
	\operatorname{var}(\widehat{\theta}_q)
	\gtrsim
	\frac{
		3\psi (N_o-1)\lambda_{\rm len}^2
	}
	{
		2\pi^2 W^2 (N_o+1)
		|\rho_q|^2
		\sin^2\theta_q
	} .
	\label{eq:th_scalar_aoa_crb}
\end{equation}
Equation \eqref{eq:th_scalar_aoa_crb} gives a direct interpretation: AoA estimation becomes more accurate when the path SNR $|\rho_q|^2/\psi$, the aperture $W$, or the number of ports $N_o$ increases, and becomes more difficult when the steering vector is less sensitive to angular perturbations.

\subsubsection{Channel-Domain CRB Induced by Joint Parameter Estimation}

The CRB in \eqref{eq:th_joint_crb_xi} is defined for the parameter vector $\boldsymbol{\xi}$. To obtain the corresponding channel-domain CRB, define the real-equivalent active channel vector as $
\bar{\mathbf{x}}_{\mathcal{S}}
\triangleq
\left[
\Re\{\mathbf{x}_{\mathcal{S}}\}^{\mathrm{T}},
\Im\{\mathbf{x}_{\mathcal{S}}\}^{\mathrm{T}}
\right]^{\mathrm{T}}
\in
\mathbb{R}^{2K_aN_o}$. Let $\mathbf{t}_q(\theta_q)\in\mathbb{C}^{K_aN_o}$ denote the channel-domain basis vector that places $\mathbf{u}(\theta_q)$ in the block corresponding to user $\mathsf{k}(q)$ and zeros elsewhere. Then
$
\mathbf{x}_{\mathcal{S}}
=
\sum_{q=1}^{L_{\Sigma}}
\rho_q\mathbf{t}_q(\theta_q)$. Furthermore, define $
\mathbf{T}(\boldsymbol{\theta})
\triangleq
\left[
\mathbf{t}_1(\theta_1),
\ldots,
\mathbf{t}_{L_{\Sigma}}(\theta_{L_{\Sigma}})
\right]$,
and $
\mathbf{D}_x(\boldsymbol{\rho},\boldsymbol{\theta})
\triangleq
\left[
\rho_1\dot{\mathbf{t}}_1(\theta_1),
\ldots,
\rho_{L_{\Sigma}}\dot{\mathbf{t}}_{L_{\Sigma}}(\theta_{L_{\Sigma}})
\right]$.
The complex Jacobian of $\mathbf{x}_{\mathcal{S}}$ with respect to $\boldsymbol{\xi}$ is
\begin{equation}
	\mathbf{G}_{x,c}(\boldsymbol{\xi})
	=
	\left[
	\mathbf{T}(\boldsymbol{\theta}),
	\,
	j\mathbf{T}(\boldsymbol{\theta}),
	\,
	\mathbf{D}_x(\boldsymbol{\rho},\boldsymbol{\theta})
	\right].
	\label{eq:th_complex_channel_jacobian}
\end{equation}
Therefore, the real Jacobian of $\bar{\mathbf{x}}_{\mathcal{S}}$ is
\begin{equation}
	\mathbf{G}_{x}(\boldsymbol{\xi})
	=
	\frac{\partial \bar{\mathbf{x}}_{\mathcal{S}}}{\partial \boldsymbol{\xi}}
	=
	\begin{bmatrix}
		\Re\{\mathbf{T}\}
		&
		-\Im\{\mathbf{T}\}
		&
		\Re\{\mathbf{D}_x\}
		\\
		\Im\{\mathbf{T}\}
		&
		\Re\{\mathbf{T}\}
		&
		\Im\{\mathbf{D}_x\}
	\end{bmatrix}.
	\label{eq:th_real_channel_jacobian}
\end{equation}
By the CRB for differentiable functions of parameters,
\begin{equation}
	\operatorname{Cov}
	\left(
	\widehat{\bar{\mathbf{x}}}_{\mathcal{S}}
	-
	\bar{\mathbf{x}}_{\mathcal{S}}
	\right)
	\succeq
	\mathbf{G}_{x}(\boldsymbol{\xi})
	\mathbf{J}_D^{-1}(\boldsymbol{\xi})
	\mathbf{G}_{x}^{\mathrm{T}}(\boldsymbol{\xi}).
	\label{eq:th_channel_domain_cov_crb}
\end{equation}
Thus, the channel-domain MSE satisfies
\begin{equation}
	\mathbb{E}
	\left[
	\left\|
	\widehat{\mathbf{x}}_{\mathcal{S}}
	-
	\mathbf{x}_{\mathcal{S}}
	\right\|_2^2
	\bigm|
	\boldsymbol{\xi}
	\right]
	\ge
	\operatorname{tr}
	\left[
	\mathbf{G}_{x}(\boldsymbol{\xi})
	\mathbf{J}_D^{-1}(\boldsymbol{\xi})
	\mathbf{G}_{x}^{\mathrm{T}}(\boldsymbol{\xi})
	\right],
	\label{eq:th_joint_channel_crb}
\end{equation}
which is the joint gain-AoA channel-domain CRB. {\em This bound is the most relevant deterministic benchmark for the proposed algorithm, since the algorithm estimates both the path gains and the AoAs.} All CRB expressions above are MSE bounds and the corresponding NMSE bounds can be obtained by normalizing them with the active-channel energy. In summary, the analytical benchmarks are summarized in Table~\ref{tab:Summary and Comparison} for quick reference.

\begin{table*}[t!]
	\centering
	\caption{Summary and Comparison of Derived Estimation Limit Bounds}
	\label{tab:Summary and Comparison}
	\begingroup
	\footnotesize
	\renewcommand{\arraystretch}{1.15}
	\setlength{\tabcolsep}{4pt}
	\setlength{\arrayrulewidth}{0.35pt}
	\begin{threeparttable}
		\begin{tabular*}{\textwidth}{@{\extracolsep{\fill}}ll|c|c|c|c|c|c|c@{}}
			\Xhline{0.8pt}
			\multicolumn{2}{c|}{\multirow{2}{*}{MSE Bound Type}}
			& \multirow{2}{*}{Signal Model}
			& \multicolumn{4}{c|}{Side Information}
			& \multirow{2}{*}{Expression}
			& \multirow{2}{*}{\makecell{Performance\\Comparison}} \\
			\cline{4-7}
			\multicolumn{2}{c|}{}
			&
			& Activity
			& Angle
			& Path Num.
			& Path Strength
			&
			& \\
			\hline
			Unstructured
			& Achievable
			& \eqref{eq:2}
			& \checkmark
			& \ding{55}
			& \ding{55}
			& \ding{55}
			& \eqref{eq:th_unstructured_oracle_crlb}
			& Worst \\
			\hline
			\multirow{2}{*}{Geometry-Structured}
			& Achievable
			& \multirow{2}{*}{\eqref{eq:th_structured_linear_model}}
			& \checkmark
			& \ding{55}
			& \checkmark
			& \ding{55}
			& \eqref{eq:th_joint_channel_crb}
			& Medium \\
			\cline{2-2}\cline{4-9}
			& Lowest-Bound
			&
			& \checkmark
			& \checkmark
			& \checkmark
			& \ding{55}
			& \eqref{eq:th_structured_oracle_crlb}
			& Best \\
			\Xhline{0.8pt}
		\end{tabular*}
		\begin{tablenotes}[flushleft]
			\footnotesize
			\item \emph{Note:} $\checkmark$ and \ding{55} indicate that the corresponding side information is known and unknown, respectively. If the path strength is assumed to be known a priori, the remaining channel information can be directly derived. Therefore, the all-known case is not separately compared.
		\end{tablenotes}
	\end{threeparttable}
	\endgroup
	\vspace{-6mm}
\end{table*}

\begin{table}[t]
	\centering
	\caption{Common configurations and simulation settings.}
	\label{tab:parameters}
	\begingroup
	\footnotesize
	\renewcommand{\arraystretch}{1.12}
	\setlength{\tabcolsep}{3pt}
	\setlength{\arrayrulewidth}{0.35pt}
	
	\begin{tabularx}{\columnwidth}{@{}>{\raggedright\arraybackslash}p{0.34\columnwidth}|>{\raggedright\arraybackslash}X@{}}
		\Xhline{0.8pt}
		\multicolumn{2}{@{}c@{}}{\textbf{Common Configurations}} \\
		\hline
		Angle and distance
		& $d\in[20,100]$ m, $\theta\in[30^\circ,150^\circ]$ \\
		\hline
		Baseline algorithm
		& $T_{\max}=50$, early-stop residual $10^{-5}$ \\
		\hline
		Proposed algorithm
		& \makecell[l]{
			$\Omega_{\theta}=30^\circ\!:\!1^\circ\!:\!150^\circ$, $L_{\max}=3$, $\kappa=0.15$;\\
			$\lambda_{\min}=0.02$, $C_{\max}=2$;
			$K_{c,\max}=5K_a$,\\ $\phi_{\min}=10^{-10}$, $\phi_{\max}=10^{2}$;
		} \\
		\hline
		Others
		& \makecell[l]{
			$K=1000$, $f(d)=d^{-2}$; $K_r=3$, $L_s=3$;\\
			$M=64$
		} \\
		\Xhline{0.8pt}
	\end{tabularx}
	
	\vspace{1mm}
	
	\begin{tabularx}{\columnwidth}{@{}>{\raggedright\arraybackslash}X|>{\centering\arraybackslash}p{0.18\columnwidth}|>{\centering\arraybackslash}p{0.08\columnwidth}|>{\centering\arraybackslash}p{0.15\columnwidth}|>{\centering\arraybackslash}p{0.07\columnwidth}@{}}
		\Xhline{0.8pt}
		\multicolumn{5}{@{}c@{}}{\textbf{Figure-Specific Settings}} \\
		\hline
		Figure & SNR & $G$ & $N_o$ & $K_a$ \\
		\hline
		Fig.~\ref{fig:Performance_comparison}: Algorithms comparison (NLOS, $K_r=0$)
		& x-axis & 150 & 16 & 50 \\
		\hline
		Fig.~\ref{fig:Performance_vs_PathNum}: Path number $L_{\max}$
		& 5 dB & 100 & $\{8,16,32\}$ & 25 \\
		\hline
		Fig.~\ref{fig:Convergence}: Convergence behavior
		& $\{-10,10\}$ dB & 100 & 16 & 50 \\
		\hline
		Fig.~\ref{fig:Performance_vs_SamplingRatio}: Subsampling ratio impact
		& 0 dB & x-axis & 16 & 50 \\
		\hline
		Fig.~\ref{fig:Performance_vs_SNR}: Performance versus SNR
		& x-axis & 100 & $\{8,32\}$ & 50 \\
		\hline
		Fig.~\ref{fig:Performance_vs_Antenna}: Performance versus observation number $N_o$
		& $\{-10,10\}$ dB & 100 & x-axis & 50 \\
		\hline
		Fig.~\ref{fig:Complexity_per_Iteration}: Complexity comparison
		& -- & 100 & 16 & 50 \\
		\Xhline{0.8pt}
	\end{tabularx}
	
	\endgroup
	\vspace{-3mm}
\end{table}

\section{Numerical Results}\label{sec.numerical_results}

In this section, we validate the derived estimation bounds and the proposed algorithmic framework through extensive simulations. Unless otherwise specified, the system configurations are listed in Table~\ref{tab:parameters} and remain unchanged throughout this section. For the large-scale fading coefficient (LSFC), we adopt the commonly used model $f(d)=d^{-2}$. The performance metrics include the activity detection error (ADE), channel normalized mean square error (NMSE), and AoA NMSE, which are defined as $
		\mathrm{ADE} = 1-\frac{\lvert \mathcal{S}\cap \widehat{\mathcal{S}} \rvert}{K_a},~
		\mathrm{NMSE}_{\mathrm{ch}} = 
		\frac{\mathbb{E}\left[\|\mathbf{h}_k-\widehat{\mathbf{h}}_k\|_2^2\right]}
		{\mathbb{E}\left[\|\mathbf{h}_k\|_2^2\right]},~
		\mathrm{NMSE}_{\mathrm{AoA}} = 
		\frac{\mathbb{E}\left[\|\boldsymbol{\theta}_k-\widehat{\boldsymbol{\theta}}_k\|_2^2\right]}
		{\mathbb{E}\left[\|\boldsymbol{\theta}_k\|_2^2\right]}$.
Here, $\mathcal{S}$, $\mathbf{h}_k$, and $\boldsymbol{\theta}_k$ denote the true active-user set, channel vector, and AoA vector containing $L_s$ angles in radians for the $k$-th active user, respectively. Their corresponding estimates are denoted by $\widehat{\mathcal{S}}$, $\widehat{\mathbf{h}}_k$, and $\widehat{\boldsymbol{\theta}}_k$. The following evaluation conventions are adopted:
\begin{itemize}
	\item[-] For activity detection, simultaneous orthogonal matching pursuit (SOMP) requires prior knowledge of the number of active users and outputs exactly $K_a$ candidate active users.
	\item[-] For fair comparison and unified ADE evaluation, in all AMP-based algorithms, the $K_a$ codewords with the largest activity likelihoods $\lambda_k$ are declared active.
	\item[-] The channel and AoA NMSEs are averaged only over correctly detected active users.
	\item[-] When $L_{\max}>L_s$, i.e., the assumed maximum number of paths exceeds the true number, only the $L_s$ angle estimates associated with the largest estimated path strengths are used for AoA NMSE calculation. If fewer than $L_s$ angles are returned, the estimated AoA vector is zero-padded.
\end{itemize}

Moreover, the \textit{received} SNR is defined as
\begin{equation}\label{eq:29}
	\mathrm{SNR}
	=\frac{\|\mathbf{a}_k\|_2^2\mathbb{E}\left[\|\mathbf{h}_k\|_2^2\right]}
	{\mathbb{E}\left[\|\mathbf{Z}\|_{\mathsf{F}}^2\right]}
	=\frac{\bar{\varsigma}}{\psi G N_o},
\end{equation}
where $\|\mathbf{a}_k\|_2^2=1$ follows from the unit-power pilot assumption in Section~\ref{sec.channel model}, and $\bar{\varsigma}\triangleq \frac{1}{K_a}\sum_{k=1}^{K_a}\varsigma_k$ denotes the average LSFC of all active users. Since the user locations are randomly generated in each Monte Carlo trial, the AWGN variance is adjusted in each realization to maintain the target received SNR. The subsampling ratio of the sensing codebook is defined as $\frac{G}{K}$.

The practical algorithmic benchmarks include conventional EM-AMP~\cite{EM-AMP1} in Algorithm~\ref{alg:algorithm1}, SOMP+LS~\cite{CE5}, the angle-codebook-based method~\cite{FAS_channel 0.2}, AMP with angular refinement~\cite{FAS_amp_jstsp}, and AMP exploiting geographical side information~\cite{FAS_amp_jstsp}. The theoretical benchmarks, averaged over Monte Carlo trials, are denoted as follows:
\begin{itemize}
	\item[-] {\em Bound I (B.I)}: the unstructured bound summarized in Table~\ref{tab:Summary and Comparison} and expressed in \eqref{eq:th_unstructured_oracle_crlb}. It characterizes the expected performance without exploiting additional geometrical diversity.
	\item[-] {\em Bound II (B.II)}: the structured bound summarized in Table~\ref{tab:Summary and Comparison} and expressed in \eqref{eq:th_joint_channel_crb}. It exploits geometrical diversity while treating both the AoAs and path strengths as unknown parameters to be estimated, thereby representing the achievable limit of a practical structured design.
	\item[-] {\em Bound III (B.III)}: the structured oracle bound summarized in Table~\ref{tab:Summary and Comparison} and expressed in \eqref{eq:th_structured_oracle_crlb}. It exploits geometrical diversity with known AoA information, and thus serves as the best-case oracle benchmark.
\end{itemize}

\begin{figure}[t!]
	\centering
	\includegraphics[width=\columnwidth]{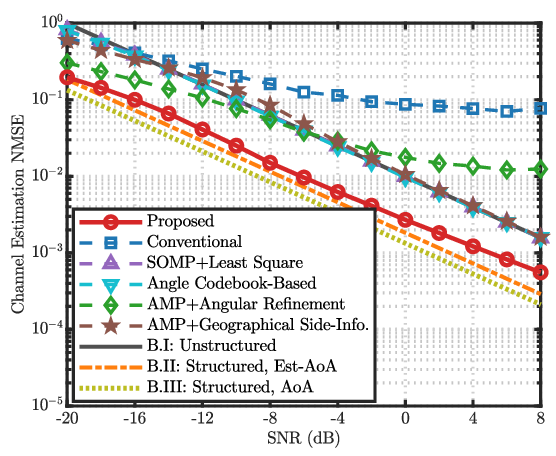}
	\caption{Performance comparison with state-of-the-art benchmarks, including conventional EM-AMP~\cite{EM-AMP1} in Algorithm~\ref{alg:algorithm1}, SOMP+LS~\cite{CE5}, the angle-codebook-based method~\cite{FAS_channel 0.2}, AMP with angular refinement~\cite{FAS_amp_jstsp}, and AMP exploiting geographical side information~\cite{FAS_amp_jstsp}, under the NLOS setting $K_r=0$, $G=150$, $N_o=16$, $K_a=50$, and $L_{\max}=L_s=3$.}
	\label{fig:Performance_comparison}
	\vspace{-4mm}
\end{figure}
\begin{figure}[t!]
	\centering
	\includegraphics[width=0.8\columnwidth]{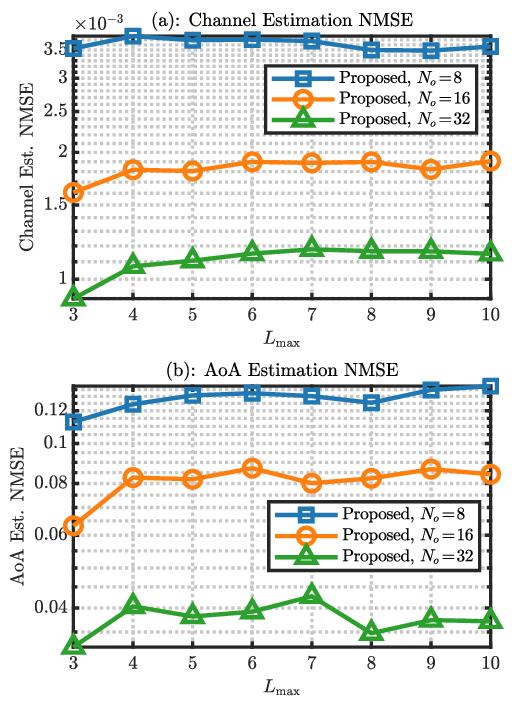}
	\caption{Impact of the assumed maximum number of paths $L_{\max}$ on the proposed algorithmic framework under $\mathrm{SNR}=5$~dB, $G=100$, $N_o\in\{8,16,32\}$, and $K_a=25$.}
	\label{fig:Performance_vs_PathNum}
	\vspace{-4mm}
\end{figure}
\begin{figure}[t!]
	\centering
	\includegraphics[width=0.8\columnwidth]{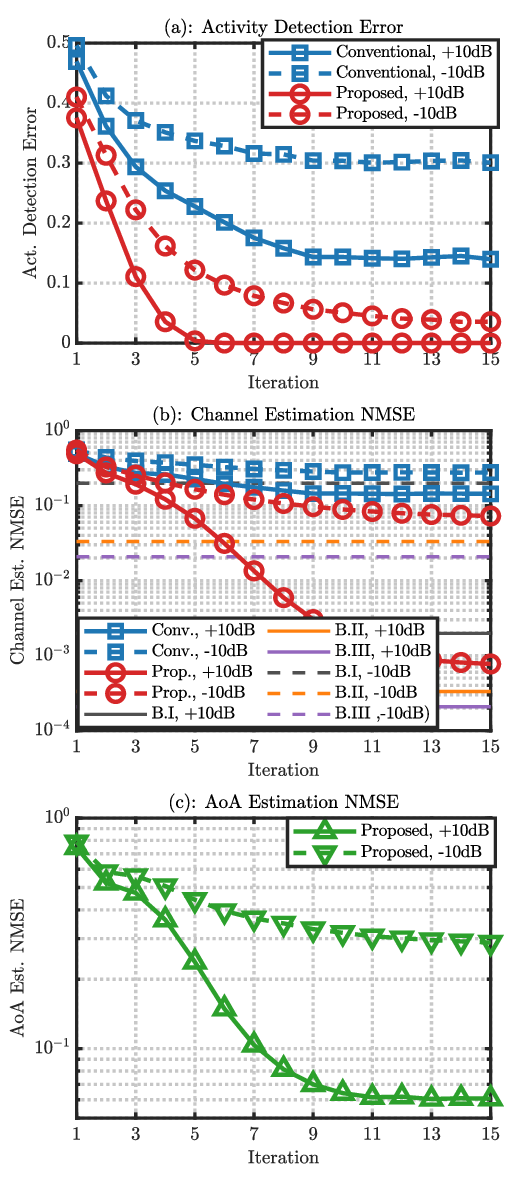}
	\caption{Convergence behavior of different algorithms under $\mathrm{SNR}\in \{-10,10\}$~dB, $G=100$, $N_o=16$, and $K_a=50$.}
	\label{fig:Convergence}
	\vspace{-4mm}
\end{figure}
\begin{figure}[t!]
	\centering
	\includegraphics[width=0.8\columnwidth]{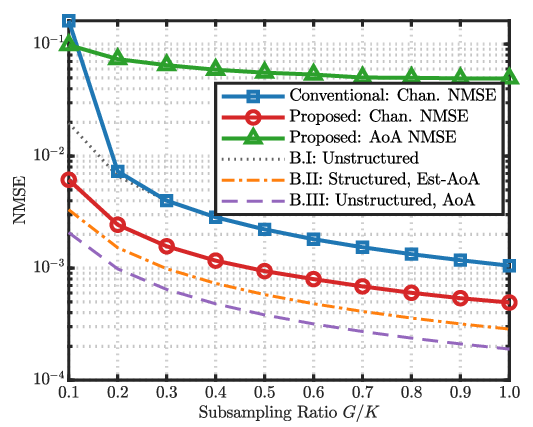}
	\caption{Channel reconstruction performance versus the subsampling ratio $G/K$, with $K=1000$ and varying pilot length $G$, under $\mathrm{SNR}=0$~dB, $N_o=16$, and $K_a=50$.}
	\label{fig:Performance_vs_SamplingRatio}
	\vspace{-4mm}
\end{figure}

In the following, the proposed algorithmic framework and the derived estimation bounds are validated through comparison with state-of-the-art benchmarks.

\paragraph{Limit-Approaching Performance}
Fig.~\ref{fig:Performance_comparison} compares the proposed framework with existing schemes and the derived bounds. Bound~I gives the loosest NMSE, while Bound~III gives the tightest one, consistent with the amount of structural information exploited in Table~\ref{tab:Summary and Comparison}. The greedy baselines, including SOMP+LS and angle-codebook refinement, together with geographical-information-aided AMP, closely follow the unstructured Bound~I, validating it as the baseline benchmark. In contrast, the proposed method tightly approaches Bound~II without AoA prior information. Fig.~\ref{fig:Performance_vs_PathNum} further shows that the proposed framework is insensitive to $L_{\max}$, since redundant path gains are automatically suppressed when $L_{\max}>L_s$. Hence, $L_{\max}=L_s$ is adopted unless otherwise specified.

\begin{figure}[t!]
	\centering
	\includegraphics[width=0.8\columnwidth]{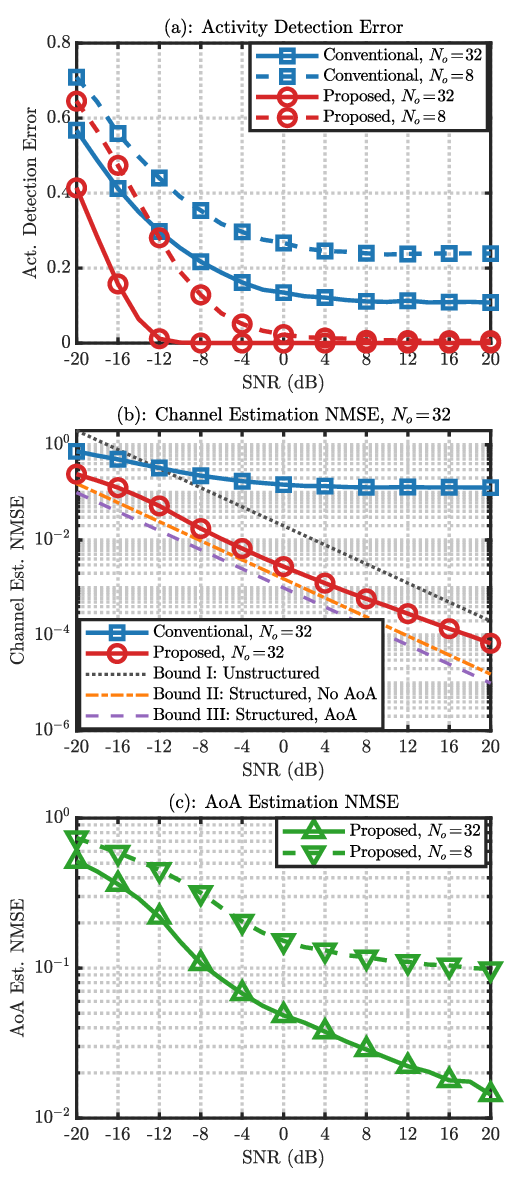}
	\caption{Detection and estimation performance versus SNR under $G=100$, $N_o\in \{8,32\}$, and $K_a=50$.}
	\label{fig:Performance_vs_SNR}
	\vspace{-4mm}
\end{figure}

\paragraph{Robustness and Convergence}
Figs.~\ref{fig:Convergence}--\ref{fig:Performance_vs_SamplingRatio} show that both the proposed method and conventional EM-AMP converge within about 10 iterations under the stopping tolerance $10^{-5}$. However, conventional EM-AMP is sensitive to the subsampling ratio and approaches Bound~I only when sufficient measurements are available. The angular-refined AMP and conventional EM-AMP also suffer from high-SNR error floors due to insufficient subsampling and steering-vector correlation. By contrast, the proposed framework remains robust to subsampling, consistently follows Bound~II, and avoids evident error floors.

\begin{figure}[t!]
	\centering
	\includegraphics[width=\columnwidth]{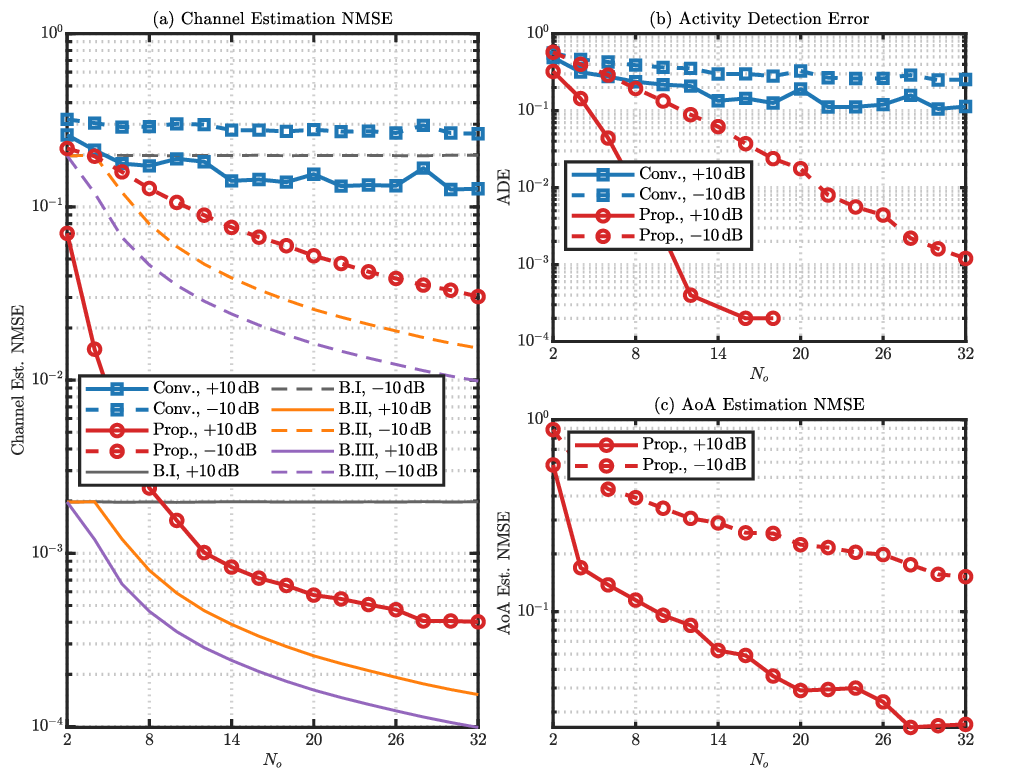}
	\caption{Performance versus $N_o$ under $\mathrm{SNR}\in \{-10,10\}$~dB, $G=100$, and $K_a=50$.}
	\label{fig:Performance_vs_Antenna}
	\vspace{-4mm}
\end{figure}

\paragraph{SNR and Observation Gains}
Fig.~\ref{fig:Performance_vs_SNR} shows that, under $\frac{G}{K}=0.1$, conventional EM-AMP saturates at high SNR, whereas the proposed algorithm continues to improve with SNR and follows the decreasing trend of the derived bounds. Fig.~\ref{fig:Performance_vs_Antenna} further shows that increasing $N_o$ improves both channel and AoA NMSEs by enhancing angular resolution, while conventional EM-AMP gains little due to residual interference.

\begin{figure}[t!]
	\centering
	\includegraphics[width=0.8\columnwidth]{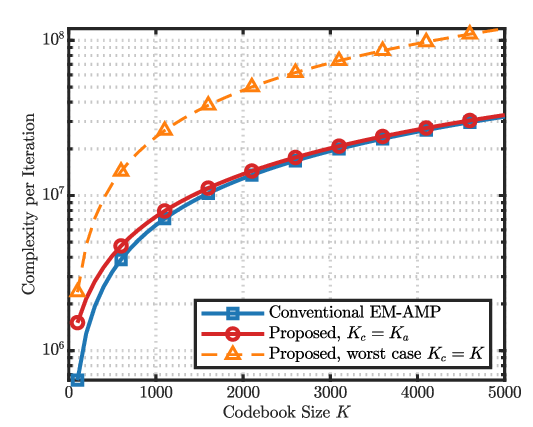}
	\caption{Per-iteration complexity under $G=100$, $N_o=16$, and $K_a=50$.}
	\label{fig:Complexity_per_Iteration}
	\vspace{-4mm}
\end{figure}

\paragraph{Complexity}
Fig.~\ref{fig:Complexity_per_Iteration} compares the per-iteration complexity. Since angular updates are performed only for active-likely codewords, the proposed method incurs only moderate overhead over conventional linear-complexity EM-AMP. Even in the worst case $K_c=K$, the complexity increase is modest; for example, at $K=1000$, it requires about twice the per-iteration complexity while achieving over two orders of magnitude lower NMSE in Fig.~\ref{fig:Performance_vs_Antenna}.

\section{Conclusion}
In conclusion, this work establishes fundamental reconstruction benchmarks and develops a robust GS-EM-AMP framework that exploits AoA-domain geometry to achieve near-limit CSI reconstruction accuracy with manageable complexity in both conventional and FASs.


\begin{thebibliography}{1}
\bibliographystyle{IEEEtran}
\smaller
\bibitem{survey1}
R. W. Heath, N. González-Prelcic, S. Rangan, W. Roh and A. M. Sayeed, ``An overview of signal processing techniques for millimeter wave MIMO systems,'' {\em IEEE J. Sel. Topics Signal Process.}, vol. 10, no. 3, pp. 436-453, April 2016.
\bibitem{Geo1}
H. Chu, L. Zheng, and X. Wang, ``Super-resolution mmWave channel estimation for generalized spatial modulation systems,'' {\em IEEE J. Sel. Topics Signal Process.}, vol. 13, no. 6, pp. 1336–1347, Oct. 2019.
\bibitem{Geo4}
Z. Zhang, J. Dang and Z. Zhang, ``Unsourced random access under quasi-static fading channel: A less-is-more strategy,'' {\em IEEE Trans. Wireless Commun.}, vol. 25, pp. 3287-3300, 2026.
\bibitem{low_rank1}
S. Liang, X. Wang, and L. Ping, ``Semi-blind detection in hybrid massive MIMO systems via low-rank matrix completion,'' {\em IEEE Trans. Wireless Commun.}, vol. 18, no. 11, pp. 5242–5254, Nov. 2019.
\bibitem{low_rank2}
Z. Zhang, J. Dang, Z. Zhang, L. Wu and B. Zhu, ``Unsourced random access with uncoupled compressive sensing and forward error correction,'' {\em IEEE Trans. Veh. Technol.}, vol. 74, no. 2, pp. 3555-3560, Feb. 2025.
\bibitem{Geo2}
J. Liu and X. Wang, ``Sparsity-exploiting blind receiver algorithms for unsourced multiple access in MIMO and massive MIMO channels,'' {\em IEEE Trans. Commun.}, vol. 69, no. 12, pp. 8055–8067, Dec. 2021.
\bibitem{Geo3}
Z. Zhang, J. Dang, Y. Qi, Z. Zhang, L. Wu and H. Wang, ``Efficient ODMA for unsourced random access in MIMO and hybrid massive MIMO,'' {\em IEEE Internet Things J.}, vol. 11, no. 23, pp. 38846-38860, 1 Dec., 2024.

\bibitem{fas-twc-21}
K. K. Wong, A. Shojaeifard, K.-F. Tong and Y. Zhang, ``Fluid antenna systems,'' \emph{IEEE Trans. Wireless Commun.}, vol. 20, no. 3, pp. 1950--1962, Mar. 2021.
\bibitem{kit_electronic}
K. K. Wong, K. F. Tong, Y. Chen, Y. Zhang, ``Closed-form expressions for spatial correlation parameters for performance analysis of fluid antenna systems,'' {\em Elect. Lett.}, vol. 58, no. 11, pp. 454--457, Apr. 2022.
\bibitem{CE5}
H. Xu, G. Zhou, K.-K. Wong, W. K. New, C. Wang, C.-B. Chae, R. Murch, S. Jin, and Y. Zhang, ``Channel estimation for FAS-assisted multiuser mmWave systems,'' {\em IEEE Commun. Lett.}, vol. 28, no. 3, pp. 632-636, March 2024.
\bibitem{FAS_channel 0.2}
K. Zhou, K. Zhou, Z. Zhang, J. Dang, Q. Sun and Z. Zhang, ``Sparsity-exploiting channel estimation for unsourced random access with fluid antenna'', in {\em Proc. IEEE Veh. Technol. Conf. (VTC2025-Spring)}, Oslo, Norway, 2025, pp. 1-7.
\bibitem{CE4}
B. Xu, Y. Chen, Q. Cui, X. Tao, K.-K. Wong ``Sparse bayesian learning-based channel estimation for fluid antenna systems,'' {\em IEEE Wirel. Commun. Lett.}, vol. 14, no. 2, pp. 325-329, Feb. 2025.
\bibitem{FAS_amp_jstsp}
Z. Zhang, J. Dang, D. Morales-Jimenez, H. Jiang, Z. Zhang, C. Masouros, and C.-B. Chae, ``Joint activity detection and channel estimation for fluid antenna system exploiting geographical and angular information,'' {\em IEEE J. Sel. Topics Signal Process.}, early access, \url{DOI: 10.1109/JSTSP.2026.3673148}.

\bibitem{KL_fas_channel_estimation}
Y. Wu, Z. Zhang, H. Jiang, K. -K. Wong and C. -B. Chae, ``Learned-approximate message passing under Karhunen–Loève modeling for fluid antenna systems,'' {\em IEEE Wireless Commun. Lett.}, vol. 15, pp. 2719-2723, 2026.
\bibitem{CE6}
C. Skouroumounis and I. Krikidis ``Fluid antenna with linear MMSE channel estimation for large-scale cellular networks,'' {\em IEEE Trans. Commun.}, vol. 71, no. 2, pp. 1112-1125, Feb. 2023.


\bibitem{EM-AMP1}
J. P. Vila and P. Schniter, ``Expectation-maximization Gaussian-mixture approximate message passing,'' {\em IEEE Trans. Signal Process.}, vol. 61, no. 19, pp. 4658-4672, Oct.1, 2013.


\end{thebibliography}
\end{document}